# Uniform definition of comparable and searchable information on the web

Wolfgang Orthuber, University Medical Center Schleswig-Holstein, Kiel, Germany

Basically information means selection within a domain (value or definition set) of possibilities. For objectifiable, comparable and precise information the domain should be the same for all. Therefore the global (online) definition of the domain is proposed here. It is advantageous to define an ordered domain, because this allows using numbers for addressing the elements and because nature is ordered in many respects. The original data can be ordered in multiple independent ways. We can define a domain with multiple independent numeric dimensions to reflect this. Because we want to search information in the domain, for quantification of similarity we define a distance function or metric. Therefore we propose "Domain Spaces" (DSs) which are online defined nestable metric spaces. Their elements are called "Domain Vectors" (DVs) and have the simple form:

URL (of common DS definition) plus sequence of numbers

At this the sequence must be given so that the mapping of numbers to the DS dimensions is clear. By help of appropriate software DVs can be represented e.g. as words and numbers. Compared to words, however, DVs have (as original information) important objectifiable advantages (clear definition, objectivity, information content, range, resolution, efficiency, searchability). Using DSs users can define which information they make searchable and how it is searchable. *DSs can be also used to make quantitative (numeric) data as uniform DVs interoperable, comparable and searchable.* The approach is demonstrated in an online database with search engine (http://NumericSearch.com). The search procedure is called "Numeric Search". It consists of two systematic steps:

1. Selection of the appropriate DS e.g. by conventional word based search within the DS definitions.
2. Range and/or similarity search of DVs in the selected DS.

Categories and Subject Descriptors: **C.2.6 [Internetworking]:** Standards; **H.3.3 [Information search and retrieval]:** Search process – Selection process; **E.1 [Data structures]:** Distributed data structures; **I.5.2 [Pattern recognition]:** Design Methodology, Feature evaluation and selection

Additional Key Words and Phrases: Metric Space, Domain Space, DS, Domain Vector, DV, Feature Vector, Metric Search, High Resolution Search, Numeric Search, Interoperable quantitative data Medical Documentation and Decision Support see 8.10 and 8.11.

## 1. INTRODUCTION

This new version of the article has a new generalized topic. We want to emphasize that the proposed "Domain Spaces" for interoperable searchable quantitative data can be used generally for definition of internationally well defined information. Implicitly information is regarded as something which determines a selection from a set (of possibilities). Also science uses this viewpoint. Already Kolmogorov [Kolmogorov 1968] wrote at the start of his elementary combinatorial approach:

'Assume that a variable x is capable of taking values in a finite set X
containing N elements. We say that the "entropy" of the variable x is
$H(x) = \log_2 N$
By giving x a definite value
$x = a$
we "remove" this entropy and communicate "information"
$I = \log_2 N$.'

The term "giving x a definite value" means "selection of x within the set X containing N elements". Subsequently we use the word **"domain"** for the (value or definition) set X. We know that $I = \log_2 N$ corresponds to the number of bits of the communicated information necessary for selection of x within the domain. Obviously the receiver of the information also needs to know the domain to use the communicated information. This preknowledge can be seen as "present" information necessary for exact interpretation (usage) of the communicated "new" information.[1] Preknowledge of the domain by all participants of communication is precondition for exact communication.

<u>In language based communication (e.g. here) the domain is the common vocabulary</u> (with all grammatical variations and specialist extensions).[2] In everyday life there are often relevant misunderstandings of the provided information due to different (interpretation resp.) preknowledge of the domain. Therefore here we try to find the best possible (maximally exact) solution within the current international infrastructure for communication - the web (internet). On the web we have the opportunity to define the domain <u>once</u> at <u>one</u> location which is associated with an international unique name: the "Uniform Resource Locator" or "URL" (please look at [8.16]). For communication (transfer of information) we can combine this URL (of the domain definition) with information which is necessary for exact selection of an element within the domain. It is advantageous to define an ordered domain, because this allows using numbers for addressing the elements and because nature is ordered in many respects. If the original data are ordered in multiple independent ways, we can define a domain with multiple independent numeric dimensions to reflect this, so that every element is addressable by a sequence of numbers. Because we want to search information (resp. perform similarity search) in the domain, for quantification of similarity we can define a distance function or metric. Therefore it is advantageous to define every domain as multidimensional metric space. On the web many domains can be defined and existing definitions can be reused. Therefore as domains we propose "Domain Spaces" (DSs) which are on online defined nestable metric spaces. The URL of the DS definition is used as identifier of the DS. With this searchable information can be represented in simple form as "Domain Vectors" (DVs):

<u>URL (of DS definition) **plus** a sequence of numbers</u> [8.16]

At this the sequence must be given so that the mapping of numbers to the DS dimensions is clear. By help of appropriate software DVs can be illustrated and represented e.g. as words and numbers. Compared to words, however, DVs have (as original information) important objectifiable advantages (clear definition, objectivity, information content, range, resolution, efficiency, searchability). Using DSs users can define which information they make searchable and how it is searchable. DSs can be used for globally uniform definition of complex information, e.g. medical original data and findings (http://numericsearch.com/FutureMedPoster.pdf).

At this the quantitative (ordered) structure of DSs can be used to get maximal resolution and precision. DSs can make quantitative data searchable in a way that their elements, the DVs, can be viewed like words of a common language in a common dictionary - a click on it can quickly show its content. With this all well defined quantitative data can become searchable and viewable [8.13] like elements of the common vocabulary - obviously a huge vocabulary. Precise reality conform (original) information is usually quantitative. Results of feature extraction (of complex information, e.g. sounds, pictures, patterns), measurements, precise descriptions of products and other things, medical findings, technical data, scientific data, economic data, commercial data - all these are examples for quantitative (numeric) data. Therefore quantitative data are important. But quantitative data are up to now not searchable on the web (which is important also for grouping, classification and analysis), and the analysis of huge amounts of data [Big Data] is very complicated due to missing identification of data and inconsistent data structures. Therefore this paper also provides a systematic approach to more structured data on the web.

---

[1] It is interesting that this order of "present" to "new" corresponds to the physical interpretation of "time" and it would be interesting to deepen this within development of an information theoretical approach towards physics of spacetime. But focus of this paper is efficient communication on the web.

[2] Implicitly it is assumed that participants of communication know the definition of the domain (common vocabulary). But even if exactly the same native language is spoken, different participants of communication cannot have an exactly identical definition of the domain up to the physical level. We see that full physical exactness of communication (which includes exact spacetime coordinates) is out of range in this macroscopic (non-quantum physical) surrounding alone due to time delay.

The potential of quantitative description is large but up to not nearly used, because they are not part of the common vocabulary. They are machine-recognizable usually only within isolated applications and not generally interoperable. This, however, is desirable in many areas, e.g. in medicine. Therefore the approach guides users that they can publish own structured machine readable definitions of quantitative data. Interoperable reusage and combination of existing definitions within new (complex) definitions is part of the concept. Data are provided structured according to existing definitions and identified, so they are machine readable and interoperable. The search window is automatically structured according to the selected definition of the user.

Conventional word based search cannot reach the precision of this, because words usually represent a rough categorization of measurable reality [Black et al. 1963], [Holsgrove et al. 1998]. A finer description is possible using words in combination with numbers (quantitative data) [Nakao et al. 1983]. Example: Compared to the word "cupboard" the term "cupboard, price = 250 Euro, width = 100 cm, height = 200 cm, depth = 50 cm" contains additional information which frequently is decision relevant and therefore important. At this every number represents a quantity [Wikipedia: Quantity 2014] which is a property that can exist as a magnitude or multitude. Quantities can be compared in terms of "more", "less" or "equal", or by assigning a numerical value in terms of a unit of measurement. The above string "width = 100 cm, height = 200 cm, depth = 50 cm" describes quantitative data (the size) of a cupboard. It becomes clear, that searching quantitative data means search of well defined numbers. Because human brain is usually adapted to words of language within their context numbers are seldom in everyday language and the importance of quantitative (numeric) data is usually underestimated. But this is only a subjective impression, from the objective point of view quantitative data are very important as original information. Original (natural) data are ordered. All results of physical measurements are quantitative (numeric) data. These form not only the unbiased basis of our perception, also derived information, e.g. technical data, results of feature extraction (most relevant information about a certain object) [Wikipedia: Feature extraction 2014] can be most efficiently represented in well defined numeric (quantitative) form. Numeric (quantitative) data are often important and decision relevant. But up to now these data are not searchable. This results from the fact that up to now quantitative data (numbers) are on the web usually missing or given in non-uniform way, using heterogeneous units and definitions. There are approaches to extract quantitative data from existing web content, especially from tables [Sarawagi et al. 2014] [Pimplikar et al. 2012] [Limaye et al. 2010]. The results are not error free but they can help at conversion of unsharp (without clear rule given) quantitative web data into well defined machine readable form. Today for this web data can be stored according to the Linked Data approach of the semantic web [Scientific American: The semantic web 2001]. OWL [McGuinness 2004] can be used for numeric definitions. In RDF [World Wide Web Consortium 2014] these definitions can be addressed using URLs. These approaches are comprehensive and started already before the year 2000. Later more slender syntax proposals for structured data have been introduced. An important initiative is the microdata approach [WHATWG 2014] used in https://schema.org/ [schema.org 2014]. It allows integrating structured data directly into web pages using HTML tags. But all realized approaches together have not been sufficient.

**Despite enormous activities concerning search and the semantic web up to now quantitative data are not searchable on the web.** This is a clear shortcoming because it can be solved. It seems that the costs of delay are underestimated. We list some desirable features of

a possible solution: An interesting and practicable approach to searchability (and machine readability) of quantitative (numeric) data should have the following features:

a) It is general (globally usable for all quantitative data).

b) It is slender (avoid unnecessary overhead at the basis).

c) It is hierarchic (to allow high complexity by nesting).

d) Quantitative data are searchable by similarity (user can determine the order of the search result by providing "wished" numbers).

e) Quantitative data are searchable without special knowledge (after providing a keyword or *topic* search of quantitative data should be possible by filling in a form).

f) Data providers get clear rules, e.g. authors of websites can provide quantitative data with the aid of adapted software by filling a form, which is e.g. determined by an online definition (of the used DS as described below).

g) Quantitative data of a web resource are visible after click on it (e.g. in tabular form, using adapted web browsers).

[schema.org 2014] contains a vocabulary for a set of items and its hierarchic structure implies clearness. But it is no general approach a). For this it is necessary that the users of the search engine can globally define the searchable numeric data. Among the above features up to now a) d) e) f) g) are not realized. Condition e) is important for practicability, we cannot expect that users know a priori a large vocabulary or names of searchable variables (e.g. to use these in SPARQL queries).

The here shown approach makes quantitative data searchable and is designed to fulfill above conditions a) b) c) d) e) f) g). This can be already demonstrated using the online implementation http://NumericSearch.com [Orthuber 2012].

As already outlined above, quantitative data are represented as "Domain Vectors" (DVs) (identified number sequences) which are elements of "Domain Spaces" (DSs). Every DS represents a worldwide defined metric space [Zezula et al. 2005] with unique Domain Space Identifier (DSI) which is the URL of the DS definition. The DS has a finite count of dimensions (Fig. 1) and it is nestable, i.e. every dimension can represent a (unbranched) value (usually a number, or text if explicitly specified) or again a DS (Fig. 2). The DS can be defined by any domain name owner according to a certain domain [Haas 2005] of interest. The DVs are the elements of a DS. Every DV provides identified quantitative data and so well defined similarity relations to other DVs of a DS. For this we assume that an unbranched dimension of a DS (a value, see Fig. 2) is numeric. An unbranched value can be also a text, but this case is here not used for similarity comparison but for additional textual conditions in search, see 6.8.

**Domain Space (DS):**

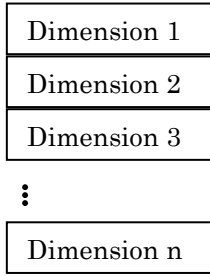

Fig. 1. A multidimensional DS. The DS and every of its dimensions have a unique name (URL).

**Dimension of a DS:**

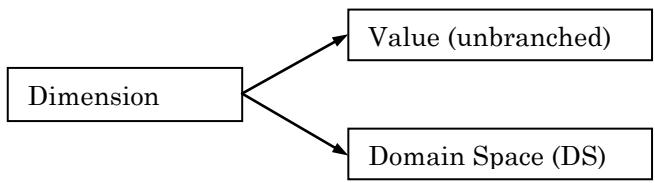

Fig. 2. A dimension of a DS can represent an unbranched value or a DS. An unbranched value is usually a number, but in an extended approach (if explicitly specified) it can be also text (e.g. a hyperlink) or binary data. If a dimension represents a value, it is called unbranched, else (if the dimension represents a DS) it is called branched. Because a dimension (of a DS) again can represent a DS, DS definitions can be nested and reused within new DS definitions.

The structure of DSs definitions (Fig. 1 and Fig. 2) is comparable to that of an ordered directory tree: A DS combines dimensions which represent values or again DSs, like a directory contains files or again directories. At this in a DS definition the order of dimensions is relevant because it determines the order in which values can be given in a DV without necessary dimension identifier. Additionally in a DS circular definitions (8.7.1) are possible.

Representation of objects and data by such online (worldwide) defined vectors (DVs) can become a fundamental concept of informatics due to its basal features. It is a precise, internationally interoperable, comparable and searchable representation of information. The paper explains details of the approach. It is organized as follows: Section 2 recounts the metric space concept and its application, especially in case of partially defined vectors. Section 3 describes the Minkowski distance function and its adaptability. Based on this nesting of distance functions is derived which allows integration and combination of metric space definitions. Section 4 expands on the DS concept and its potential for connection of information. Section 5 addresses reusage of DS definitions within new definitions. Section 6 demonstrates the concept by examples using the online implementation. Section 6.7 provides details about the necessary content of DS definitions and DVs for development of a web standard. Section 8 discusses several important aspects. The conclusion follows in section 9.

## 2. METRIC SPACES

Metric spaces are a natural container for searchable quantitative data, because all elements (vectors) of a metric space have a well defined distance which can be used for similarity calculation. There is a huge literature on search in metric spaces, also the high dimensional case is well investigated. Approximate nearest neighbor search is possible in sub linear time also in high dimensional spaces [Indyk et al. 1998]. Here we also use the metric space concept. On the web, however, we usually cannot assume a fixed set of dimensions for search ([4.4](#)), therefore we introduced the dimension-wise synchronized index ([6.5](#)) which are suitable for all combinations of dimensions.

HTTP-URLs are the basis for direct connections on the web via hyperlinks, and they can be also used for identification of (online definitions of) metric spaces which define connections (similarity relations) between *all* their elements. At this the term "similarity" is represented in well defined way by a nonnegative real number called "distance": The smaller the distance between two elements, the greater is their similarity. If the distance between two elements is zero, the compared quantitative data are identical.

Because metric spaces play a central role in this paper, their definition is repeated here: A metric space [[Wikipedia: Metric space 2014](#)] is a set S with a distance function *D* which represents for every two elements (vectors) *X,Y* in *S* the distance between X and Y as a nonnegative real number *D(X,Y)* with

$D(X,Y)=0$ if and only if *X=Y* and $\quad\quad\quad\quad\quad\quad\quad\quad\quad\quad\quad\quad\quad\quad$ (1)

$D(X,Y)+D(Y,Z) \geq D(X,Z)$ (triangle inequality) and $\quad\quad\quad\quad\quad\quad\quad$ (2)

$D(X,Y)=D(Y,X)$ $\quad\quad\quad\quad\quad\quad\quad\quad\quad\quad\quad\quad\quad\quad\quad\quad\quad\quad$ (3)

A distance function which fulfils [(1)(2)(3)](#) is called a *metric*. The distance D quantifies the similarity in S. Two elements *X,Y* of *S* are the more similar, the smaller the distance *D(X,Y)* is. Note that *D(X,Y)* is a real number and therefore one-dimensional. But the set *S* can be a multidimensional space. In this paper (as content of a DS it is a m-dimensional feature space which is subset of $\mathbf{R}^m$, and the elements (*X,Y,Z* in the above formulas) are m-dimensional feature vectors which are represented by sequences of real numbers ($x_1 \ldots x_m, y_1 \ldots y_m, z_1 \ldots z_m$). The definition of S does not contain a limitation regarding cardinality, so metric spaces can be very large.

## 2.1 Induced Metric

If $D(X,Y)$ fulfils [(1)(2)(3)](#) for all $X,Y \in S_m \subseteq \mathbf{R}^m$, then it is possible to compare a subset of all dimensions of $\mathbf{R}^m$ using an induced metric:

Let $J=\{x_{j1}, x_{j2},\ldots, x_{jn}\} \subseteq \{x_1,x_2,\ldots,x_m\}$ denote a selected subset of dimensions. We define the set $\mathbf{R}_J{}^m = \{(x_1, x_2,\ldots, x_m)$ where ($x_j \in \mathbf{R}$) for $x_j \in J$ and (($x_j \in \mathbf{R}$ or $x_j$ is undefined ) for $x_j \notin J$ }

So in contrast to $\mathbf{R}^m$ in $\mathbf{R}_J{}^m$ values at dimensions outside J can be undefined.

Let $X_m$, $Y_m \in \mathbf{R}^m$ with $X_m = (x_1, x_2,\ldots, x_m)$ and $Y_m = (y_1, y_2,\ldots, y_m)$. We define the mapping

$B_J: \mathbf{R}_J{}^m \rightarrow S_J \subseteq \mathbf{R}^m$ with $B_J(X_m)=(b_1, b_2,\ldots, b_m)$

$\quad$ where $b_j=x_j$ for $x_j \in J$ and $b_j=0$ for $x_j \notin J$ $\quad\quad\quad\quad\quad\quad\quad\quad\quad\quad$ (4)

So $B_J$ simply replaces possibly undefined values by 0, therefore $B_J(X_m)$ is well defined. Its value set $S_J$ is a subspace of $\mathbf{R}^m$ [[Wikipedia: Linear subspace 2014](#)]. So if $D_m: \mathbf{R}^m \times \mathbf{R}^m \rightarrow \mathbf{R}$ is a metric on $\mathbf{R}^m$, then the restriction $D_J: S_J \times S_J \rightarrow \mathbf{R}$ with

$D_J(X_m,Y_m)=D_m(X_m,Y_m)$ $\quad\quad\quad\quad\quad\quad\quad\quad\quad\quad\quad\quad\quad\quad\quad$ (5)

is a metric on $S_J$ . It is called induced metric. The subspace $S_J$ forms together with $D_J$ a metric space.

## 2.2 Comparable vectors

In applications we cannot expect that $X_m$ contains values at all dimensions $x_1, x_2,\ldots, x_m$ of $\mathbf{R}^m$ . But we can expect, that $X_m$ contains values at an adaptable subset $J$ of dimensions, i.e. we can assume $X_m \in \mathbf{R}_J{}^m$ and $B_J(X_m) \in S_J$ according to [(4)](#). In this case $X_m$ is called **comparable in $S_J$**.

Two vectors are called **comparable**, if they have values at an overlapping set of dimensions. The distance between comparable vectors depends on the selected subset J of compared dimensions. These determine the space $S_J$ in which the distance $D_J$ (5) is calculated.

## 3. SIMILARITY COMPARISON OF QUANTITATIVE DATA

Similarity comparison is done by calculation of the distance between the vectors (Domain Vectors) which represent the compared data in their space (Domain Space).

### 3.1 The Distance Function

Generally for similarity comparison every metric (see 2) can be used as distance function. If the triangle inequality (2) is not needed, it is even not necessary that the distance function is a metric [Aggarwal et al. 2001]. The optimal distance function depends on the application, and on the definition of "optimal".

### 3.2 The Minkowski distance

Because it is not possible to discuss every distance function, we need to make a first preselection (which can be expanded later). Because the dimensionality of Domain Spaces can vary, we need a distance function with adaptable dimensionality. Nesting should be possible (see 3.5). Frequently used distance functions like Euclidean and Manhattan distance should be included as special cases. The Minkowski distance [Wikipedia: Minkowski distance 2014] covers these requirements and is established.

The Minkowski distance D(X,Y) of order $k \geq 1$ between two vectors X = ($x_1$, $x_2$,..., $x_n$) and Y = ($y_1$, $y_2$,..., $y_n$) $\in$ $\mathbf{R}^n$ is

$$D(X,Y) = \left( \sum_{j=1}^{n} \left| x_j - y_j \right|^k \right)^{\frac{1}{k}} ; \qquad (k \geq 1) \tag{6}$$

Here we presuppose $k \geq 1$, because in this case the Minkowski distance D fulfills besides (1)(3) also (2) and is a metric. In (6) there is freedom regarding the unit or scale of $x_j$, $y_j$, therefore we can multiply every dimension with a constant $r_j > 0$ and get the weighted Minkowski distance.

### 3.3 The weighted Minkowski distance

The (with constants $r_j > 0$) weighted Minkowski distance D(X,Y) of order $k \geq 1$ between two vectors X = ($x_1$, $x_2$,..., $x_n$) and Y = ($y_1$, $y_2$,..., $y_n$) $\in$ $\mathbf{R}^n$ is

$$D(X,Y) = \left( \sum_{j=1}^{n} (r_j \left| x_j - y_j \right|)^k \right)^{\frac{1}{k}} \tag{7}$$

D is a metric for $k \geq 1$ and we can search given quantitative data $x_1$, $x_2$,..., $x_n$ by inserting the $x_j$ with $r_j > 0$ as coordinates of a searched feature vector $X_0$ as described in 2.2 and calculating and sorting the distances to $X_0$. At this the factor $r_j$ determines the relative weight of dimension j. The larger $r_j$, the more dimension j influences the distance and derived search results. In case of $r_j=0$ the dimension $x_j$ is ignored. If we calculate (7) within a subspace $S_J$ (4), it is sufficient to sum up only over dimensions in J (therefore an index should require disk access only for the searched dimensions, see 6.5.).

### 3.4 Selection of the exponent k

Due to their importance we explicitly write weighted Minkowski distances with certain k. Special cases are the weighted Manhattan distance with k=1, the weighted Euclidean distance with k=2 and the weighted Maximum distance with k→∞:

$$D_1(X,Y) = \sum_{j=1}^{n} \left| r_j (x_j - y_j) \right| \qquad \text{Manhattan distance} \qquad (8)$$

$$D_2(X,Y) = \sqrt{\sum_{j=1}^{n} \left| r_j (x_j - y_j) \right|^2} \qquad \text{Euclidean distance} \qquad (9)$$

$$D_\infty(X,Y) = \max_{j=1}^{n} \left| r_j (x_j - y_j) \right| \qquad \text{Maximum distance} \qquad (10)$$

Among these $D_1$ provides the best contrast between different vectors [Aggarwal et al. 2001]. $D_2$ can be used e.g. for calculating distances with direct geometrical meaning. $D_\infty$ can be e.g. used for limiting the range of dimensions.

### 3.5 Nested distance functions

It is efficient to use definitions of established metric spaces within other new definitions. For this combination of distance functions to one nested distance function is required. We now show, that analogously to (7) instead of differences |x_j-y_j| also metrics can be nested to a superordinated metric:

PROPOSITION.    Let V denote a vector space whose dimensions are a concatenation of the dimensions of vector spaces $V_1$, $V_2$,..., $V_n$ and $X,Y,Z \in V$. We presuppose $w_j > 0$, $k \geq 1$ and that for j $\in$ {1,2,...,n} $D_j$ is a metric on $V_j$ and $X_j, Y_j, Z_j \in V_j$. Then the following *nested* distance function is a metric:

$$DC(X,Y) = \left( \sum_{j=1}^{n} \left( w_j D_j (X_j, Y_j) \right)^k \right)^{\frac{1}{k}} \qquad (11)$$

PROOF.    We have to show (1)(2)(3) of section 2. Due to nonnegativity of $D_j(X_j,Y_j)$ from DC(X,Y)=0 follows $D_j(X_j,Y_j)$=0 and therefore $X_j$=$Y_j$ for j$\in${1,2,...,n} and X=Y. Reversely from X=Y follows $X_j$=$Y_j$ and $D_j(X_j,Y_j)$=0 for j $\in$ {1,2,...,n} and so DC(X,Y)=0, therefore (1) is true. Symmetry (3) of DC(X,Y) follows from symmetry of $D_j(X_j,Y_j)$. We now prove (2). For k$\geq$1 and $u_j$, $v_j$ $\in$ **R** we have due to the Minkowski inequality [MATH41002 2014] :

$$\left( \sum_{j=1}^{n} \left| u_j + v_j \right|^k \right)^{\frac{1}{k}} \leq \left( \sum_{j=1}^{n} \left| u_j \right|^k \right)^{\frac{1}{k}} + \left( \sum_{j=1}^{n} \left| v_j \right|^k \right)^{\frac{1}{k}}$$

we set $u_j$=$w_j D_j(X_j,Y_j)\geq$0 and $v_j$=$w_j D_j(Y_j,Z_j)$ $\geq$0 and get

$$\left( \sum_{j=1}^{n} \left( w_j (D_j (X_j, Y_j) + D_j (Y_j, Z_j)) \right)^k \right)^{\frac{1}{k}} \leq$$

$$\left( \sum_{j=1}^{n} \left( w_j D_j (X_j, Y_j) \right)^k \right)^{\frac{1}{k}} + \left( \sum_{j=1}^{n} \left( w_j D_j (Y_j, Z_j) \right)^k \right)^{\frac{1}{k}}$$

Due to $D_j(X_j,Z_j) \leq D_j(X_j,Y_j)+D_j(Y_j,Z_j)$ (triangle inequality)

$$\left( \sum_{j=1}^{n} \left( w_j D_j (X_j Z_j) \right)^k \right)^{\frac{1}{k}} \leq$$

$$\left( \sum_{j=1}^{n} \left( w_j D_j (X_j, Y_j) \right)^k \right)^{\frac{1}{k}} + \left( \sum_{j=1}^{n} \left( w_j D_j (Y_j, Z_j) \right)^k \right)^{\frac{1}{k}}$$

this is just the triangle inequality for DC in [(11)](#):
DC(X,Z) ≤ DC(X,Y) + DC(Y,Z)

So we have shown also [(2)](#) and so proven that [(11)](#) is a metric. [(11)](#) has a similar structure like the Minkowski metric [(7)](#), only the $|x_j-y_j|$ are replaced by $D_j(X_j,Y_j)$. Because the $|x_j-y_j|$ are special cases of a metric, [(11)](#) is a generalization of [(7)](#). So in a (weighted) Minkowski distance function [(7)](#) for every j the absolute difference $|x_j-y_j|$ can be replaced by a metric $D_j(X_j,Y_j)$. The result [(11)](#) remains a metric.

Important special cases of [(11)](#) are the following nested distance functions

$$DC_1(X,Y) = \sum_{j=1}^{n} \left( w_j D_j (X_j, Y_j) \right) \qquad \text{nested Manhattan distance} \qquad (12)$$

$$DC_2(X,Y) = \sqrt{\sum_{j=1}^{n} \left( w_j D_j (X_j, Y_j) \right)^2} \qquad \text{nested Euclidean distance} \qquad (13)$$

$$DC_\infty(X,Y) = \max_{j=1}^{n} \left( w_j D_j (X_j, Y_j) \right) \qquad \text{nested Maximum distance} \qquad (14)$$

### 3.6 Estimation of the weights

We assume that the functions $D_j(X_j,Y_j)$ in [(11)](#) are Minkowski distances of the form [(7)](#). The weights $r_j$ and $w_j$ are free parameters and we need an estimation. If there is no further information available, the initial values are $r_j=1$ in [(7)](#) and $w_j=1$ in [(11)](#). The $w_j$ can be left unchanged, because all necessary modifications can be done by adjusting the $r_j$ in [(7)](#). They are important for multidimensional similarity search, because they determine the relative weight of dimensions for calculation of the overall distance.

If there are no individual preferences, at least the influence of a dimension's unit should be eliminated. The smaller the unit of a dimension, the larger is its numerical variation. It is possible to take into consideration the numerical variation of a dimension j by setting $r_j = 1/s_j$, where $s_j$ can be e.g. the standard deviation (or a difference between two given percentiles) of dimension j. Calculation of the $r_j$ can be done also retroactively.

We now come to the application of the above theoretical background.

## 4. DOMAIN SPACES

The metric space concept can be used systematically on the web to make multidimensional quantitative data searchable. When defining a metric space about a certain domain of interest, regularly it is desirable to include all possibly interesting dimensions. This leads to a domain specific metric space $DS_m$ with maximal dimensionality m. Because we cannot expect that users always provide values for all m dimensions we introduce the following convention:

*A Domain Space (short "DS") is defined by the domain specific metric space $DS_m$ with maximal dimensionality $m$. The elements of a DS are called Domain Vectors . Every Domain Vector (short "DV") has a feature vector which has values at all or a subset of all $m$ dimensions. Partially defined feature vectors can be mapped by [(4)](#) into subspaces of $DS_m$.*

Subsequently we abbreviate, where clear, the term "$DS_m$ of DS" by "DS". For example we call "subspaces of $DS_m$ of the DS" simply "subspaces of the DS". Analogously we abbreviate, where clear, "feature vector of the DV" simply by "DV".

DVs are called "comparable" if their feature vectors are comparable (see [2.2](#)), i.e. if they have values at an overlapping set of dimensions. It is possible to select a subset J of these dimensions for similarity comparison and to calculate distances $D_J$ [(5)](#) between all DVs which are comparable in $S_J$ [(4)](#). The smaller the distance is, the greater is the similarity in $S_J$.

The subspaces $S_J$ of a DS form a set of domain specific metric spaces [[Kriegel et al. 2010](#)], [[Wikipedia: Vector space model 2014](#)] or conceptual spaces [[Gaerdenfors 2000; 2004](#)].

The structure of a DS is shown in ([Fig. 1](#)) and ([Fig. 2](#)). To ensure that also DSs with non numeric dimensions are metric spaces we can define for every non-numeric dimension $j : |x_j - y_j| = 0$ for $x_j = y_j$ and $|x_j - y_j| = 1$ for $x_j \neq y_j$ (discrete metric) and we see that [(6)](#) and [(7)](#) fulfill the conditions [(1)(2)(3)](#) for a metric.

### 4.1 The addressable web extended by Domain Spaces

Every DS has a URL [[Berners-Lee et al. 1998](#)]. It is the URL of its definition on the web and it is the worldwide unique Domain Space Identifier (DSI). The DS consists of its definition and of its elements which are the DVs. Every DV contains at least the following information:

- Its "Vector Location" (short "VL"). This is the URL of the DV on the web.

- The DSI (URL of the DS definition)

- The numeric representation of its feature vector, given by (hyperlinks to) identified numeric values for all or a part $x_{j1}, x_{j2},\ldots, x_{jn}$ of all dimensions $x_1 \ldots x_m$ of the DS (see [2.1](#)).

Domain Vectors (DVs) can be regarded as links (or mappings) into Domain Spaces. [Fig. 3](#) compares the most important characteristics of Hyperlinks and Domain Vectors. It shows that the destination of a hyperlink can be every URL of the web, while the destination of a DV can be only the URL of a DS (which represents *all* contained DVs). The DS definition is an intermediate station of bidirectional links ([Fig. 4](#)). DVs are especially useful when multiple resources with quantitative features of the same domain of interest should be connected by "similarity". The similarity depends on the (by the user) selected subset J of compared dimensions. These determine the subspace $S_J$ in which comparison is done, and these determine the distance function $D_J$ in [(5)](#). The distance is defined to all DVs which have values at least at the selected dimensions. The similarity (of the compared dimensions) is the greater, the smaller the distance is.

|  | **Hyperlink** | **Domain Vector (DV)** |
|---|---|---|
| **Usual location** | resource with URL | resource with URL |
| **Connections from location** | unidirectional to a URL on the web | via URL of the DS bidirectional similarity relations to all (locations of) comparable DVs |
| **Connections back to location** |  | given from (the locations of) all comparable DVs of the same DS |
| **Essential information (besides location)** | URL of destination | URL of DS and values $x_i$ of a subset of dimensions of the vector |
| **Optional** |  | URL of described resource |
| **Distances** |  | available to all (locations of) comparable DVs of this DS and included DSs, see section 5 |
| **Additional purpose** |  | searchable quantitative description |

Fig. 3. Comparison Hyperlink / Domain Vector. As element of a Domain Space a Domain Vector is connected.

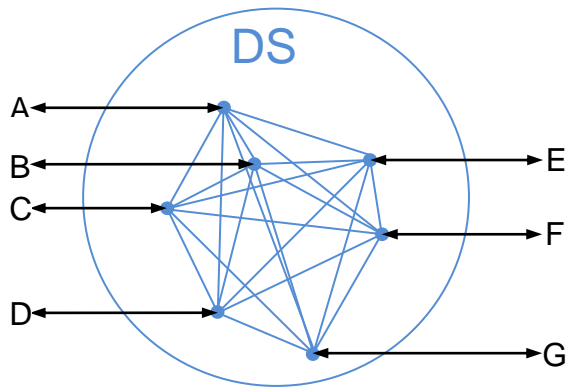

Fig. 4. Bidirectional weighted connections (similarity relations) defined by a group of DVs to a (URL of a) DS. Shown are connections within a two-dimensional subspace. They connect the (locations of the) DVs. The length of the blue lines illustrates the distance. The smaller the distance, the greater is the similarity.

As example we assume that the user has selected 2 dimensions for comparison in a DS. We can use their values as coordinates and represent every DV as point in the 2D plain. Fig. 4 shows an example with 7 comparable DVs in the subspace determined by the selected dimensions. If we assume Euclidean distance (9), the lengths of the blue lines can be used to represent the distances between the DVs .

[Fig. 5](#) shows the connections defined by a group of hyperlinks to a URL. Hyperlinks generate no implicit connections, so k hyperlinks generate also k (unidirectional) connections. When this is wished, the hyperlink is appropriate. But if a group of resources (with interesting quantitative descriptions) of the same domain should be (described and) connected, DVs are efficient. If we assume that there are k DVs in an m-dimensional DS, each with values at all m dimensions, then there are $2^m$ different subsets of dimensions which each define $k(k-1)/2$ distances (weighted connections). These can be evaluated, e.g. for similarity search. The quantitative data are also available for further calculations, e.g. statistics.

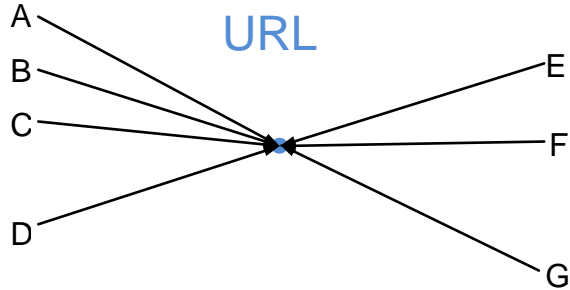

Fig. 5. Unidirectional connections defined by group of hyperlinks to the same URL. There is no bidirectional connection or similarity relation, but the destination of the hyperlinks can be every URL on the web.

### 4.2 Similarity search in a DS

In [2.2](#) is described how to calculate distances between DVs after selection of the dimensions for comparison. Analogously it is possible to select a set J of dimensions for similarity search. If $S_J$ is the subspace [(4)](#) which is determined by these dimensions, distances $D_J$ [(5)](#) can be calculated to all DVs which are comparable in $S_J$. If there are no further restrictions, these DVs form the search result. The smaller the distance of a DV, the higher is its rank in the search result. Further restrictions of the search result are possible, e.g. ranges (minima and maxima) of independently selected dimensions.

### 4.3 Storing relevant dimensions

When storing information about the world on the web, usually most measurable quantitative features are omitted. Macroscopic world has so many quantitative features, that it is necessary to select in dependence of the domain few most interesting (decision relevant) features for further information processing. These need not be "direct" physical measurement results, they can be completely derived, e.g. a result after feature extraction and/or prepared for usage with special software. For example dimensions could represent instead of or additionally to (large amounts of) original data results of specific hashing functions [[Andoni et al. 2008](#)] [[Indyk P. 1999](#)] [[Gionis et al. 1999](#)] or other effective dimension reduction techniques [[Grant et al. 2013](#)] on the original data, to increase efficiency of search. The main issue is that they are interesting (e.g. for search, or directly decision relevant) for future readers or users. So DSs typically contain those dimensions which describe the part of the world, which is interesting in the chosen domain. In the course of time new features can become interesting, other features can become deprecated. The distance function is alterable, dimensions can become unused, new dimensions can be added. This can lead to high dimensional DSs. But this does not mean that search becomes high dimensional.

### 4.4 Searching in a subset of dimensions

Due to the curse of dimensionality [[Aggarwal et al. 2001](#)] high dimensional similarity search tends to become inefficient in case of independent dimensions, else (in case of dependent dimensions) dimension reduction techniques are recommendable, see 4.3. Moreover DSs often have dimensions which describe incommensurable data, e.g. data with incommensurable units. In this case the relative weight of a dimension depends on the intention of the user at search time and it is not possible to anticipate it as described in 3.6. Therefore it is recommendable to select only a small subset J of dimensions with meaningful common distance function for similarity search (2.2) to get a well interpretable ranking of the search result. If only one dimension is included into similarity search, the search result is simply ordered by the absolute difference of the searched dimension. The smaller it is, the higher is the ranking of a DV in the search result.

Additionally the search result can be restricted by determining minima and maxima of dimensions. These dimensions are also called "searched", together with the dimensions in J.

Search is done over DVs which contain values at all searched dimensions. This implies, that the probability to be found is the greater, the more numerical values (dimensions) are given in a DV.

### 5. COMBINING DSS

### 5.1 Grouping DSs

A DS group is a set which contains as elements DSs and/or DS groups. Grouping of DSs or DS groups can be for example useful if they have a common topic. So a DS group with topic "clothes" may contain the DSs or DS groups "trousers", "shirts", "coats" etc.
Grouping of definitions can be appropriate to (simply) build a thematic structure of DSs and/or DS groups without involvement of dimension definitions.

### 5.2 Nesting DS definitions

Every DS combines an expandable set of named dimensions, and every dimension can represent:

- an unbranched value, usually a number (with selectable precision, date included, accessible to similarity search and min max conditions) or a text, if explicitly specified. Short text (tux) can be handled as efficiently as a number.

- a DS (DVs are also accessible to similarity search and min max conditions, an additional condition can be a maximal distance of its DVs.)

So a dimension can also represent another DS. It can be efficient to use DS definitions within other new definitions. For example a DS with the DSI "http://example.org/hemogram-1.htm" can be used within many DSs which describe medical findings. It is not necessary to reinvent it. We can use its metric in a nested distance function (11). This is important due to the following reasons:

- Established DS-definitions can be reused and included as "Sub-DS definitions" into new (higher dimensional) DS-definitions with weighted Minkowski metric (7), in which coordinate differences $|x_j\text{-}y_j|$ are replaced by the distances $D_j(X_j,Y_j)$ of the Sub-DSs. It is sufficient to use their URLs as reference. So also updates are automatically forwarded.
- The included DS can have any metric, also a non-Minkowski metric, or again a nested metric (11).

There are well defined similarity relations (Fig. 4) between all DVs which have a common URL of (their DS or) an included DS. One application of nested DS definitions is realization of ontology based structures [Wikipedia: Domain Ontologies 2014] in quantitative data. The *nesting level of a DS* is the maximal count of nested layers. Fig. 6 shows an example of a nested DS definition with 2 layers (nesting level 2).

Nesting of DS definitions can quickly lead to high dimensional DSs. But according to 4.4 it is recommendable to select only a small subset of dimensions for search.

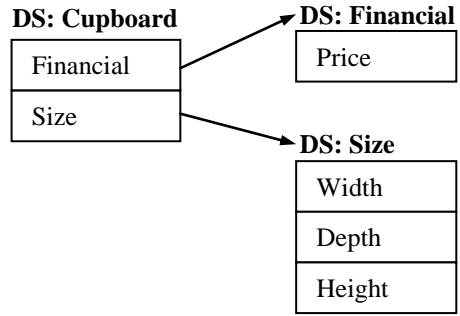

Fig. 6. Exemplary nested DS definition in the structure of Fig. 8 (nesting level is 2).

## 6. LOCAL IMPLEMENTATION

Up to now there is no web standard for worldwide valid DSs on the web. Nevertheless it is possible to implement the search principle locally. We have done this in the online implementation http://NumericSearch.com/ . It contains a local database with DSs definitions and to every DS definition a local database with the DVs of this DS.

### 6.1 Keycomments

For description of DSs, DS dimensions and DVs we used "*keycomments*" which are structured comments: Every keycomment starts with an ordered list of one or more (optionally linked) keywords, followed by a comment which is a string.

Fig. 7. Input mask of a keycomment with 3 keywords, in which the second is also a hyperlink

The advantage of the structured keycomment is that the meaning of the ordered keywords or hyperlinks can be defined a posteriori, depending on application. So for example the first keyword "kw0" can be a unique identifier, the second keyword can be a unit etc..

### 6.2 Implemented DS and DV structure

Every DS has a distance function of the form (12) or (13) or (14) where every $D_j$ has the form (8) or (9) or (10) or GPS distance (see e.g. [Movable Type Scripts 2014]). So there are exactly 2 nested layers (nesting level 2): Every DV has a vector which contains one or several "subvectors" (DVs of "Sub-DSs") with numeric dimensions. Fig. 8 shows an example of a DS with 2 subvectors. Fig. 9 shows the definition of a subvector, Fig. 10 the definition of a dimension.

## Definition of DS 1006 (Cupboard)    owner    download-def

|< << < > >> >|   0..1

| 0 | Finances |
| 0 | Price \| Euro |
| 1 | Size |
| 0 | Width \| cm |
| 1 | Depth \| cm |
| 2 | Height \| cm |

Keyword:                              Link:

| Cupboard | http://en.wikipedia.org/wiki/cupboard |
| Schrank | http://de.wikipedia.org/wiki/schrank |

Comment:

This is:  ○ draft    ○ ok    ○ deprecated

conn:  ○ Manhattan    ○ Euclidean    ○ Maximum

Fig. 8. Exemplary definition of a DS. Here the local DSI (in the local database) is "Cupboard" (in a web standard it would be a URL). The first column shows the internal index of the subvectors and the second column the internal index of the numeric dimensions (2 layers, see Fig. 6).

## Keycomment of Sub Vector    owner    vieworg

|< << < > >> >|   0..2

| 1 | Size |
| 0 | Width \| cm |
| 1 | Depth \| cm |
| 2 | Height \| cm |

Keyword:                              Link:

| Size | |

Comment:

Weight: | 1 |

Metric:   ○ GPS    ○ Manhattan    ○ Euclidean    ○ Maximum

Fig. 9. Definition of a subvector shown after clicking on "1" in the first column of Fig. 8.

Fig. 10. Definition of a dimension shown after clicking on "0" in the second column of Fig. 9.

It is possible to define minimum, maximum and weight ($r_j$ default in (7)) of every dimension. Also the representation can be adapted to the user's needs. The internal representation is a 64 bit double, the external representation can be:

**list**: Appropriate if the dimension represents the position in a list of items, e.g. a selection between two possibilities like "yes" and "no". By default the items are internally represented by integers (0,1,2...) in this implementation.
For a later implementation the following option is possible: Every item can be considered as interval, so that all items represent an ordered list of labeled intervals (a partition) of the set of real numbers R. The first interval may have no lower border, else by default the lower border of an interval is the upper border of the previous interval, the last interval may have no upper border. If these intervals are defined for a dimension and the user selects the input option "intervals", the ordered list of interval names is opened and the user can select a name. If the name is given as sort criterion, the mean of the interval (or the border, if only one border exists) is internally inserted into the similarity field. If the name is given as condition, the lower bound of the interval is (if existing) internally inserted into the min field and the upper bound (if existing) into the max field.

**tux**: Appropriate if the dimension represents a short alphanumeric text which can contain up to 8 lowercase letters a..z or digits 0..9. It is introduced because it can be easily remembered. Search is defined so that all DVs are found whose initial letters are at this dimension identical to the searched tux. So the initial letters should be most significant.
Compared to tux a list of intervals or the following ordered representations have the advantage that they allow similarity comparisons and further algebraic evaluation:

**date**: for representation dates in variable accuracy (highest significant numbers first, e.g. yyyy-mm-dd hh:mm:ss)

**floating point**: floating point number in variable accuracy, e.g. for measurement results

**integer**: integer number, e.g. for counts

An interesting but in the current version not implemented possibility is the definition of dimensions as computational results of other dimensions.

[Fig. 8](#) shows that Manhattan metric is chosen to connect the distances of 2 subvectors according to [(12)](#). In every subvector (e.g. [Fig. 9](#)) again Manhattan metric is chosen (to connect dimensions) according to [(8)](#) with all weights $r_j$=1, $w_j$=1 (1 is default value for weights). Therefore the distance function of $DS_m$ (with maximal dimensionality) is

$D(X,Y)= \quad |x_{Price} - y_{Price}| +$
$\qquad\qquad |x_{Width} - y_{Width}| + |x_{Depth} - y_{Depth}| + |x_{Height} - y_{Height}|$ (15)

According to [4.4](#) a subset J of these dimensions can be searched and so used for sorting the search result. Then the sum includes only the searched dimensions. If for example only a value of dimension "Price" is given for similarity search, then the distance function reduces to

$D(X,Y)= \quad |x_{Price} - y_{Price}|$ (16)

This distance function is used in the search example ([Fig. 13](#)) of section [6.3](#).

### 6.3 Search

The complete search procedure will be called Numeric Search subsequently. It consists of 2 systematic steps:

**1 of 2**: In the first step the appropriate DS is selected. This can be done by clicking on its index number directly in the list of all DSs ([Fig. 11](#)) or after word based search within the DSIs ([Fig. 12](#)) which are here (in the local database) the first keywords (kw0) of the space definitions. After selection of the appropriate DS its specific search mask appears ([Fig. 13](#)).

**2 of 2**: The second step is metric similarity search in the selected DS or a part of it. All data for this are provided in the search mask of the DS.

[Fig. 13](#) shows an example of a search mask with exemplary input. It shows the search of the cheapest cupboards (those nearest to price=0). Two checkboxes in column "g" are checked to signal the wish for graphical and statistical output of Price and Width. [Fig. 14](#) shows the resulting graph over the checked dimensions together with the search result.

**Select i7 (Domain Space)** [________] [ KW0 ]

|< <<< << < > >> >>> >|    1000..1030..1055

| i7 | s | r | |
|------|-----|---------|---|
| 1000 | 33 | 54 | space-of-spaces \|\| v965723 |
| 1001' | 25 | 9 | o ride |
| 1002' | 19 | 2 | o my-location |
| 1003' | 14 | 1 | o real-estate |
| 1004' | 0 | 0 | o car |
| 1005' | 194 | 10001 | o test-space \|\| try search 0..10: subv1, subv2 filled wi |
| 1006' | 423 | 24 | o Cupboard \| Schrank |
| 1007' | 47 | 11 | o Diode \|\| (for rectification) |
| 1008' | 145 | 1500001 | o 260dim-demo \|\| try search 0..10: subv1, subv2 filled |
| 1009' | 170 | 56 | o text-as-dimension-example \|\| dimensions (not used |
| 1010' | 18 | 1 | o cardiovascular-disease |
| 1011' | 1 | 0 | o diagnosis-1-copy-of-above-for-test |
| 1012' | 0 | 0 | o diagnosis-1-copy2 |
| 1013' | 0 | 0 | o diagnosis-1-copy3 |
| 1014' | 20 | 8 | o Screw \| Schraube |
| 1015' | 60 | 24 | o datacube-example-as-TS \|\| data like "The RDF Data |
| 1016' | 0 | 0 | o opinion-about-xx |
| 1017' | 1 | 0 | o climate-fluctuations |
| 1018' | 0 | 0 | o Meeting \| Treffen |
| 1019' | 9 | 11 | o Kugellager-Edelstahl |
| 1020' | 0 | 0 | o Help \|\| Search help (kind of help, time, location, dur |

Fig. 11: Excerpt of the start screen. The first column "i7" shows index numbers of the DSs. Clicking e.g. on "1006" opens the search mask ([Fig. 13](#)) of the DS with local DSI "Cupboard". The second column "s" shows the search count, the third column "r" the count of resources in a DS. Clicking on "o" in the next column shows the owner of a DS. Then follows the first obligatory keyword kw0 which here is the DSI (blue if HTTP-link), after "|" further optional keywords, after "||" a comment. After clicking on "kw0" text search is done over the DSIs of the DSs, the result is shown in Fig. 12.

### search KW0    cup

|<   <<   <   >   >>   >|    0..0

|  | i7 | ii |  |
|---|---|---|---|
| 1006 |  | 0 | o Cupboard | Schrank |

"cup" searched in 54 first keywords along i7
1 found which begins with the searched string

Fig. 12: Text search result after entering the first letters of the DSI in Fig. 11 and clicking on "kw0" (Keyword 0).

i7=1006, o | 13-02-09' Cupboard | Schrank

### DL search in DS 1006 (Cupboard)

|<   <<   <   >   >>   >|    0..1    search

|  | sim | | min | max | | g |
|---|---|---|---|---|---|---|
| 0 |  |  |  |  |  | Finances |
| 0 |  | 0 |  |  |  | ☑ Price | Euro |
| 1 |  |  |  |  |  | Size |
| 0 |  |  |  |  |  | ☑ Width | cm |
| 1 |  |  |  |  |  | ☐ Depth | cm |
| 2 |  |  |  |  |  | ☐ Height | cm |

o: ☐   w: ☐   pcnt: ☐

Fig. 13: Similarity search mask. It appears after selection of the DS (click on 1006 in Fig. 11 or Fig. 9). Similarity search of Price "0" is selected with graphic output of price in dependence of width.

i7=1006, o | 13-02-09° Cupboard | Schrank

**search result**  [new search]  [repeat]  [par]  search-stat  DS-stat

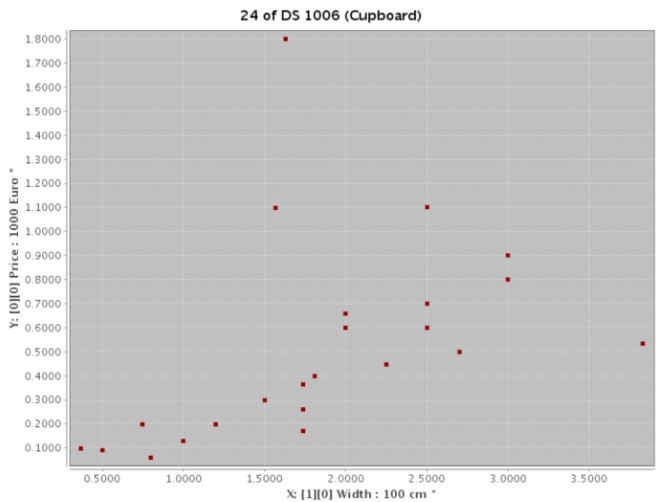

24 of DS 1006 (Cupboard)

X: av= 184.875   sd= 84.77608   min= 37   max= 383
Y: av= 515.0992   sd= 408.4305   min= 59   max= 1799.99

|< << < > >> >|    page 1

| i4 | d | a | |
|---|---|---|---|
| 9 | 59| | 0 o ikea-IVAR \| 59.00, 80, 30, 83 |
| 2 | 90| | 36 o ikea-PAX-schmal \| 90.00, 50, 60, 236 |
| 1 | 99| | 25 o ikea-PS \| 99.00, 37, 40, 190 |
| 10 | 130| | 19 o ikea-PAX \| Schrank mit 2 Tueren, weiss, Ballstad weiss \| 130.00, 100, 60, 236 |
| 5 | 170| | 3 o ikea-IVAR \| 2 Elem/ Boeden/Schrank \| 170.00, 174, 30, 179 |
| 0 | 175.4| | 3 o ikea-IVAR \| 3 Elem/Boeden/Schrank, Kiefer \| 175.40, 258, 30, 124 |
| 3 | 199| | 5 o ikea-ISALA \| 199.00, 75, 44, 131 |
| 4 | 199| | 11 o ikea-TROLLSTA \|\| Sideboard \| 199.00, 120, 50, 76 |
| 8 | 258.9| | 12 o ikea-IVAR \| 2 Elem/ Boeden/Schrank \| 258.90, 174, 50, 179 |
| 22 | 299| | 0 o Home24-Austin \|\| Schwebetuerenschrank - verschiedene Groessen - Weiss mit Mattglas \| 299.00, 150, 68, 216 |
| 6 | 362.9| | 2 o ikea-IVAR \| 2 Elem/ Schrank/Kommode \| 362.90, 174, 50, 179 |
| 11 | 399| | 14 o home24-Quadra \|\| schwebetuerenschrank, Korpus alpinweiss/Front alpinweiss \| 399.00, 181, 58, 210 |
| 12 | 449| | 4 o home24-Quadra \|\| Schwebetuerenschrank, Korpus alpinweiss/Front alpinweiss \| 449.00, 225, 58, 210 |
| 13 | 499| | 4 o home24-Quadra \|\| Schwebetuerenschrank, Korpus alpinweiss/Front alpinweiss \| 499.00, 270, 58, 210 |
| 7 | 532.7| | 17 o ikea-IVAR \| 5 Elem/ Boeden/Schrank \| 532.70, 383, 50, 226 |
| 14 | 599| | 1 o home24-Mission-4 \|\| Schwebetuerenschrank - Alpinweiss, Abs, Pearlglanz Softwhite \| 599.00, 200, 65, 218 |
| 15 | 599| | 0 o home24-Mission-4 \|\| Schwebetuerenschrank - Alpinweiss, Abs, Pearlglanz Softwhite \| 599.00, 250, 65, 218 |
| 19 | 659| | 0 o home24-Lowisville \|\| Schwebetuerenschrank - verschiedene Groessen Polarweiss/Weisslack \| 659.00, 200, 68, 216 |
| 20 | 699| | 1 o home24-Lowisville \|\| Schwebetuerenschrank - verschiedene Groessen Polarweiss/Weisslack \| 699.00, 250, 68, 216 |
| 21 | 799| | 0 o home24-Mission-4 \|\| Schwebetuerenschrank - verschiedene Groessen Polarweiss/Weisslack \| 799.00, 300, 68, 216 |
| 16 | 899| | 0 o home24-Mission-4 \|\| Schwebetuerenschrank - Alpinweiss, Abs, Pearlglanz Softwhite \| 899.00, 300, 65, 218 |
| 18 | 1099| | 1 o home24-Maxi-Eleven \|\| Kleiderschrank - verschiedene Varianten - Erle Massiv \| 1099.00, 157, 57, 203 |
| 23 | 1100| | 8 o RodrigoSchrank \| 1099.99, 250, 63, 223 |
| 17 | 1800| | 32 o home24-Rivoli \|\| Kleiderschrank - mit Dekor - Fichte, antik lackiert, Standardeinteilung \| 1799.99, 163, 60, 197 |

423 th search in DS 1006 (Cupboard) , 24 (100%) within 24 found

[new search]

Fig. 14. Search result with preceding graphic output. Click on the index in the 1. (left) column opens the DV (resource). The 2. column "d" shows the distance, here d=|price-0|, the 3. column "a" shows the access count, click on "o" in the 4. column shows data of the owner. Then follows the first keyword with optional link for description of the resource, after "||" an optional comment, after "|" the quantitative data in order of Fig. 13 (Price, Width, Depth, Height).

Together with the graph in [Fig. 14](#) statistical data of the checked dimensions are shown (average, standard deviation, minimum, maximum of price and width). This allows to check dependencies.

The min and max fields ([Fig. 13](#)) can be used to restrict the range of certain dimensions in the search result. (The checkboxes "o" and "w" can be used to restrict the search result to offered or wanted resources, the "pcnt" field allows to enter the maximal count of shown resources in the search result, pcnt=1000 is default and maximum.) So the search result can be restricted to a certain part of all DVs. Using the checkboxes of column g allows to get statistical data of selected dimensions in the search result.

According to [Fig. 13](#) the searched value of dimension "Price" is 0. This is inserted into [(16)](#) and leads to the distance d = D(X,Y) = |Price-0| = |Price|. Therefore in [Fig. 14](#) the distances in the second column "d" are equivalent to the Price of the resources which is the first number after "|". [Fig. 14](#) shows most important data of the search result in compressed form. Clicking on the index of a DV (left column) shows its data in more detailed form in a new window ([Fig. 15](#)).

i7=1006, o | 13-02-09' Cupboard | Schrank
i4=   6, o 13-02-09 ikea-IVAR || 2 Elem/ Schrank/Kommode | 362.90, 174, 50, 179

**DL 6 (Ikea-IVAR) of DS 1006** (Cupboard)   owner   vieworg   download DL

| |< << < > >> >| | 0..1 |
| 0 | Finances | |
| 0 | 362.90 | Price | Euro |
| 1 | Size | |
| 0 | 174 | Width | cm |
| 1 | 50 | Depth | cm |
| 2 | 179 | Height | cm |

o: ☑  w: ☐

Keyword:                              Link:
ikea-IVAR                            http://www.ikea.com/de/de/catalog/p
Comment:
2 Elem/ Schrank/Kommode

Fig. 15. Data of a resource shown after clicking on "6" in the left column of the table in [Fig. 14](#)

**6.4 Search results in case of equally distributed pseudo random numbers**

To illustrate the effect of the distance function [(7)](#) we generated a high dimensional Domain Space with 1500001 DVs whose dimensions have been filled with equally distributed pseudo random numbers between 0 and 10.

[Fig. 16](#) shows the search result of [(7)](#) with default $r_j$=1 in case of k=2 (Euclidean distance [(9)](#)), [Fig. 17](#) shows it in case of k=1 (Manhattan distance [(8)](#)).

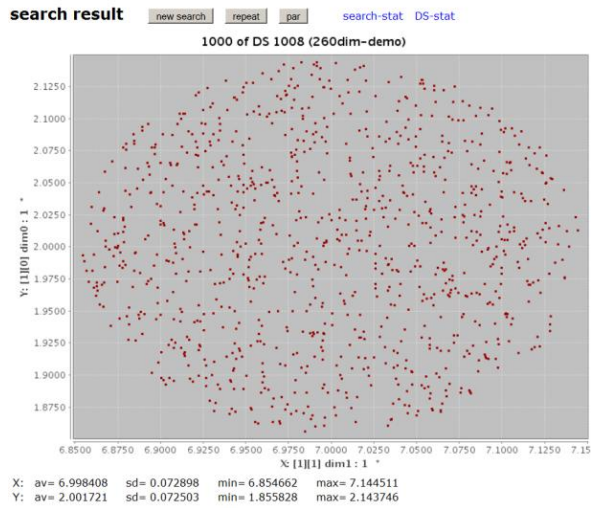

Fig. 16. Elliptic shape of output after searching within 1500001 DVs the 1000 nearest around point (x,y)=(7,2) in case of Euclidean distance [(9)](#).

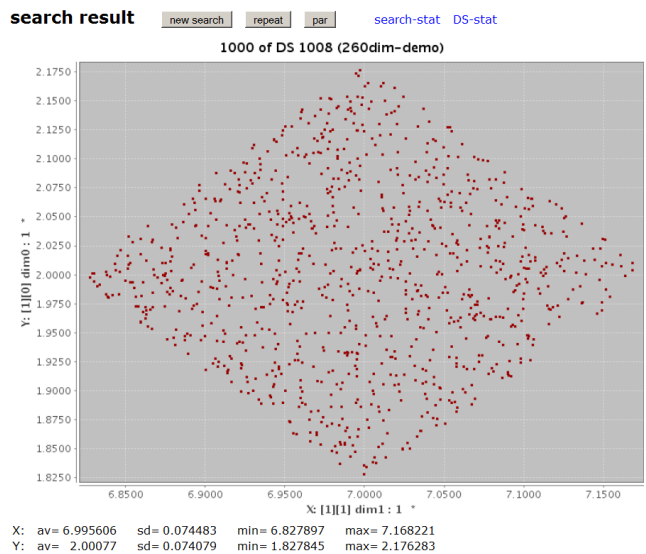

Fig. 17: Shape of output after searching within 1500001 DVs the 1000 nearest around point (x,y)=(7,2) in case of Manhattan distance [(8)](#).

While both graphs are identical near the center, in case of large deviation of one dimension (near to the border of the graph) Manhattan distance restricts the other dimension sharper than Euclidean distance. Also research [[Aggarwal et al. 2001](#)] shows that Manhattan distance provides more contrast. Therefore we selected Manhattan distance [(8)](#) as default metric. Independently of this Euclidean metric can be recommendable e.g. if the dimensions represent Cartesian coordinates and the distance should have a geometric meaning.

**6.5 Synchronized index**

In the implementation we used for every DS a *synchronized index*. Subsequently we give a short description of it in a version which can be used also for DV-groups (see 7.3) on the web.

The dimensionality of a DS (or even all DSs on the web) can become large, nevertheless similarity search usually includes only a subset of all dimensions. There are $2^n$ possibilities to select a subset of dimensions within an n-dimensional space. This selection is done before search, not before index calculation. So the synchronized index does not anticipate a certain combination of dimensions, i.e. every dimension has an own *dimension database* (three in Fig. 18) in the index which is optimized for quick access. During index creation the original DVs are scanned on the web and the count of scanned DVs (resp. DV-groups, if DVs are grouped, see 7.3) is increasing. We will call this increasing count "c". As long as dimensions belong to the same DV or DV group 7.3, the count c is constant. After a DV is fully scanned, c is stored in a separate database **(A)** together with the HTTP-URL of the DV and all further information which should be available in the search result. All (short) data records of the dimension databases get c as identifier (c in Fig. 18) of the DV and the numerical value (x4, x8 and x15 in Fig. 18) of the dimension. Data records with the same c belong to the same (multidimensional) vector of a DV. *After* all data of a DV are stored in the index, c is incremented by 1. We can say that the increasing c "synchronizes" the dimension databases **(B)**. Therefore we call this (**(A)** and **(B)**) a *synchronized index*.

So in case of multidimensional similarity search only the (few) dimension databases of the searched dimensions are serially scanned along increasing identifier c (linear performance, or better if c makes large jumps). As soon as an identifier c is found for which all searched dimension databases contain values (in Fig. 18 for c ∈ {9, 21, 29, 42}), it is checked whether the values fulfill the requirements (especially the min-max conditions). If yes, the distance is calculated ((12) can be used for combining multiple DSs). It is appended to the preliminary search result together with the dimension values and the most important information **(A)** about the DV.

This can be used for scanning all combinations of synchronized dimensions in linear performance or better. As soon as the searched dimensions are fully scanned, the preliminary search result is complete. After sorting it according to distance (the smallest distance first) we get the final search result.

Additionally special indices can be calculated, e.g. for every dimension (e.g. along x4, x8, x15 of Fig. 18) an in ascending order sorted dimension database. Such databases can be used to get quickly a maximal set which fulfils all min-max conditions. This set can be (after sort along c) used to define great jumps along c in Fig. 18 to increase scanning velocity.

If frequently only a certain dimension is searched, it can be efficient to calculate additionally an index which is *completely* sorted along this dimension so that similarity search of this dimension is possible in logarithmic time, e.g. using binary search. It is also possible (but less frequently efficient) to calculate indices which are optimized for certain *combinations* of dimensions. Calculation of additional indices for selected (combinations of) dimensions needs additional computational cost. Future research can show, under which conditions it is efficient to calculate additionally indices which are a priori optimized (e.g. by adapted sorting) for certain (combinations of) searched dimensions.

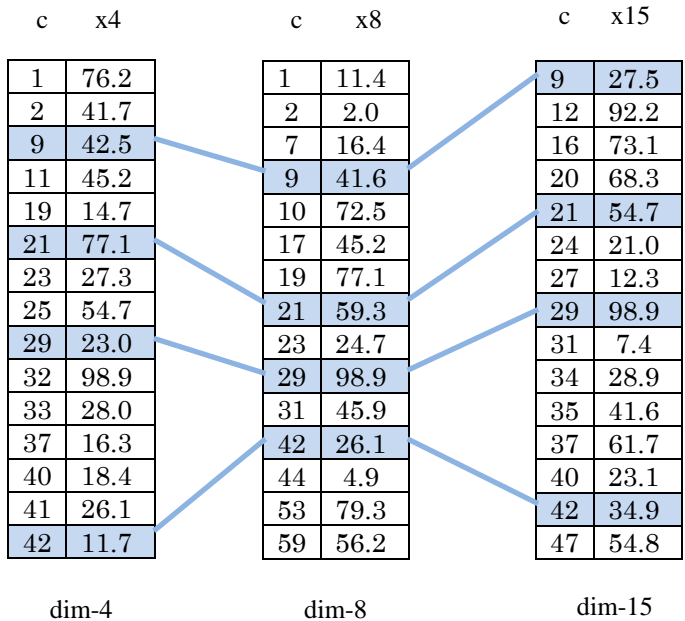

Fig. 18. Synchronized index of a DS with 3 dimension databases. Every (short) database record contains 2 numbers: The "DV number" c represents during index creation the current count of scanned DVs, x4, x8, x15 represent the numeric value of the dimension in this DV. The blue lines connect records with values from the same DV, 4 DVs are found which contain values at all 3 dimensions. Because c is increasing, the connecting blue lines cannot cross and it is possible to scan all dimensions in one pass without redundancy.

## 6.6 Index performance

The synchronized index was realized for our local database and we measured the search time. The time of multidimensional search depends not on the total dimensionality of the DS but on the dimensionality of the search (the count of simultaneously sought dimensions). Fig. 19 displays the similarity search time (for searching the most similar 1000 DVs, sorted along distance) within a 260 dimensional DS with 1500001 DVs (see 6.4). For each dimensionality 1..10 the average search time of 20 searches is shown.

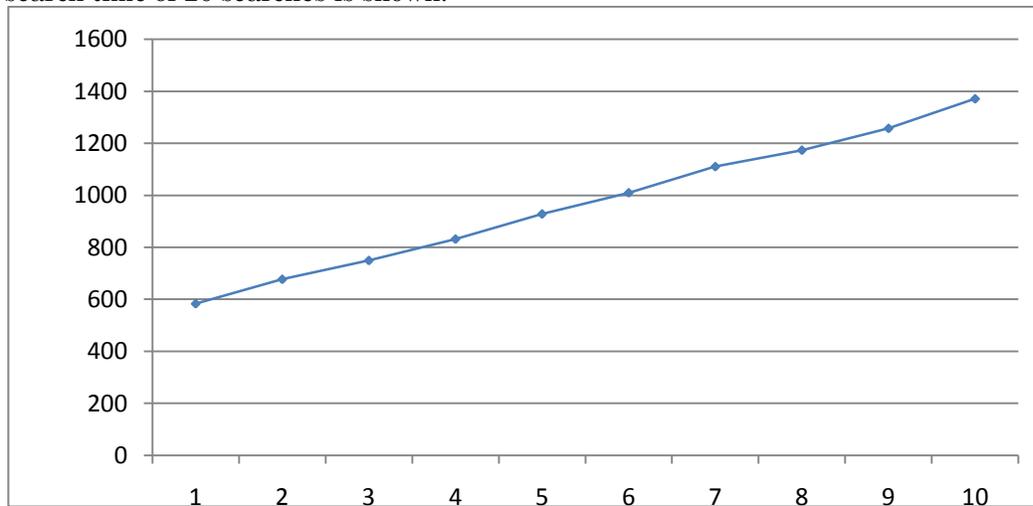

Fig. 19. Average search time in milliseconds (vertical axis) in dependence of the dimensionality of the search (horizontal axis) within a DS with 1500001 DVs.

The time for every pass of index calculation for this DS was between 23 and 24 minutes. This implementation was programmed using java jdk-7u4-linux-x64.rpm, apache-tomcat-7.0.27.zip, performance was measured on a dedicated server with Intel Core i7 3930K (3,20 GHz), 64 GB RAM, 256 GB SSD, with Linux Cent OS 6.3. Parallel processing would be possible by splitting the space and reunion of the search results.

### 6.7 Index columns for "long" dimensions which represent more information than usual

If there is a primary (not branched) dimension, it is efficient (and therefore should be usual and preferred practice) to use for it a binary representation with no more bits than the bus width of the processor, e.g. 64 bit. If the dimension has to represent more information, e.g. a text part larger than tux (e.g. a URL), it can be most efficient to distribute this information over several "conventional" dimensions (of *one* special DS, see Fig. 2), whose bits are combined to a long number which is indexed as usually by a *single* column of the synchronized index, see Fig. 18 .

### 6.8 Additional index for dimensions which represent longer text

It is possible to combine numeric search (in numeric dimensions) with additional text search in textual dimensions. If such a dimension represents text with multiple parts (words) which *all* should be searchable, then an additional index is necessary, e.g. an index of the alphabetically sorted text parts (words), each text part TP associated with an increasing list of numbers c (see Fig. 18) of DVs which contain the text part TP in this dimension. If scan (Fig. 18) is done only over DVs which have these numbers c, all DVs of the search result contain the text TP in this dimension.

### 6.9 Mapping triples into DVs of DSs, synchronized index for RDF triples

A RDF triple store can be mapped into a store of DVs of DSs (and the synchronized index). Let nmax denote the count of different predicates:

- Sort triples according to subject and map them bijectively to an increasing list of numbers c (see Fig. 18), so that all triples with the same subject have the same c (like data (dimensions) of the same DV (group)).
- Map every predicate bijectively to a dimension of a DS with nmax dimensions (or a dimension of DSs whose sum of dimensions is nmax).
- Map the object bijectively to a value of the dimension. Can be also count of different objects (index within database of all objects, e.g. if these are long texts).

So every RDF triple is mapped to one dimension of one DV (with nmax dimensions) or DV group (with together nmax dimensions). The synchronized index is suitable also in case of high dimensionality nmax.

## 7. TOWARDS STANDARDIZED DSS AND DVS

First recommendations to a web standard for DSs and DVs can be given already in this paper. DS definitions can be extended a posteriori. Already defined content cannot be changed, but commented. To meet these requirements, **keycomment-pairs** can be used. Two keycomments (Fig. 7) are joined together in a keycomment-pair. The first keycomment should already initially contain a precise and complete definition as text, the second keycomment can provide structured changeable information. A keycomment-pair has 3 possible states: "draft", "ok", and "deprecated". Once the initial (default) state "draft" is left, the first keycomment is fixed. In an appropriate environment this should ensure, that already defined essential content of a DS definition is stable.

We now summarize the content of DS definitions and DVs:

## 7.1 Content of a DS definition

- String: Domain Space Identifier (DSI): The DSI is given implicitly by the URL of the DS definition. The DSI must be stable

- "sameAs": algebraic Expression which contains URLs (DSIs) of other DS or DS dimension definitions, if existent. A simple case is another URL with or without proportionality factor.

- keycomment-pair

- ID of owner (long integer) as searchable dimension

- String: Information about the distance function (e.g. "M" for usage of [(11)](#), followed by the exponent k)

- String with one or several attributes, e.g.:

  o "c" or "connected": if in case of search usually all dimensions are given (e.g. search of GPS coordinates or results of a hash function).

  o "f" or "feature extraction": if this is an efficiently searchable feature extraction of the other content of the superior DS.

  o "n" or "not designed for search": if content of this DS is usually not directly searched (e.g. if in a superior DS combined with another DS with attribute "f").

- If there is a certain relation between the dimensions (e.g. subject, predicate, object as in a LOD triple, see [8.9](#)): keycomment-pair, which describes the relation

- Sequence of dimension definitions:

### 7.1.1 Content of a dimension definition

a) String: Dimension Identifier (DI):
To ensure that the dimension has an URL, the HTML id Attribute can be used in the form id="DI", where DI is a string which is distinct from the DIs of other dimensions of the same DS or the number of this dimension in the *original order* of dimensions. The dimension definition can be regarded as definition of a one-dimensional DS. The DSI of this DS is the URL of the dimension. It is formed by the URL of the DS definition together with the DI.

b) Rank: integer, which shows the rank (order) in representation, if deviating from the *original order* of definition (which is default order). While the original order of dimension definitions remains fixed for later compatibility (with already existing DVs, see [7.2](#)), rank can be used to define afterwards dimensions which *appear* anterior in (e.g. browser) representations and menus.

c) "sameAs": algebraic Expression which contains URLs (DSIs) of other DS or DS dimension definitions, if existent. A simple case is another URL with or without proportionality factor. Such a factor can be appropriate for measurements of the same thing by different units. After this a search engine can combine all these data in the same index column ([Fig. 18](#)).

d) keycomment-pair

e) floating point number: weight ($w_j$ in [(11)](#) if nested, else $r_j$ in [(7)](#))

f) String for content description:
   - if nested (according to section 5): DSI (URL) of the integrated DS

"virtual" dimensions (which are later decided in the DV) are not necessary, because creators of DVs can use DV groups 7.3 to complete important additional dimensions.

- if not nested: URL (DSI plus DI) of an external dimension definition
    or (if dimension is defined here):
    String which describes the content e.g.
    "integer", "floating-point- precision-8",
    "date-YYYY-MM-DD",
    "tux" or "text" (see 6.8),
    "list" etc. (see 6.2)
    borders and names of intervals, if given (option "list")
    optionally an expression which describes this
    dimension as algebraic result of other dimensions.

A DS definition can contain additional information, e.g. default preferences for representation and formatting of DVs in the Web browser. It is efficient to provide DS definitions in (for machine readability) standardized HTML, so that conventional web browsers can directly display the content in human-readable form. Similarly DVs can be embedded in HTML, with short syntax (as identified numbers), so that later people can click in their browser on a (as "Domain Vector") marked text part and a (by the DS definition determined) table opens which shows the associated quantitative data and further information about the "Domain Space" (=user defined metric space on the web).

### 7.2 Content of a DV

- String: URL of the Domain Space definition (DSI)

- If DV describes an external resource: URL of this resource

- keycomment (optional)

- ID of owner (long integer) as searchable dimension (optional)

- date: (optional) short input of date is possible without dimension definition, so a dimension definition for it is no more necessary, e.g. in the form dyyyy-mm-dd

Contents of given dimensions as semicolon separated list of numbers (this short form is possible if the original order of DIs in the DS definition is used) or a sequence of:

### 7.2.1   Content of a dimension

- String: Dimension Identifier (DI)
  The DI is optional, if the order of dimensions corresponds to the order of the dimensions in the DS definition, else the DI is necessary.

- content:
  - if dimension definition nested: integrated DV

-   or URL of integrated DV
  - if dimension definition not nested: Value (unbranched),
      depending on definition (or URL of integrated DV dimension)

- after every value its reliability can be quantified e.g. in the form "sd" where d is an estimation of the value's standard deviation

### 7.3 DV-groups

DVs can be grouped together, so that one group describes the same resource. Later it is possible to combine the dimensions of these DV-groups (i.e. dimensions of different DSs) also for search if the search engine uses a synchronized index (see 6.5) for all dimensions. This index is not restricted to one DS. *So data providers can select dimensions which they group together*, e.g. to add data which they consider as important. For a short syntax the URL of definition is sufficient for multiple dimensions as long as the data (dimension) order in the DV corresponds to the order of dimension definitions at the URL. If dimensions should be combined which are not combined at a URL, the URLs of their definitions need to be combined in the DVs.

### 7.4 Adaptable syntax: DVs as identified numbers

The concept can be realized using variable syntax. For example OWL [McGuinness 2004] and RDF [World Wide Web Consortium 2014] can be used to realize DS definitions and DVs. But because in the long run conciseness is relevant for feasibility and efficiency, we recommend to start with concise and minimal syntax with minimal overhead. Minimal precondition for DVs is that the numbers are identified, using the DSI of the DS and the DI of the dimension (see 7.1.1). Because at this the dimensions of the same DV should be grouped together, it is sufficient to specify the DSI of the DV only once at the beginning of the group. If the default order of dimensions of a DSs is inherited, it is even sufficient to use one special character as separation between dimension values in DVs (e.g. use a sequence of semicolon separated numbers).

At last we prefer as short as possible syntax. Similarly like a hyperlink a DV can contain clickable text. A possibility:

```
<v http://numericsearch.com/bw.xml; 2014-01-30; 83.914>clickable text</v>
```

The initial URL points to the DS definition, if the subsequent 2 values fit to the first 2 dimensions of the DS, no further identifiers are necessary.

## 8. DISCUSSION

Similarity search of quantitative data in metric spaces is well investigated. Due to the potential of this technique it is desirable to enhance the web by metric spaces. A DS represents a nestable metric space with unique identifier (HTTP-URL) on the web. Because DSs can be defined by web users according to their domain of interest, and because of their applicability in Domain Ontologies (see [Haas 2005]) we called these "Domain Spaces". The following subsections discuss some important aspects (partially derived from [Orthuber 2013]) of the concept:

### 8.1 Resolution and precision of DV based description and search

Due to its basal relevance the following fact is given first:

For a word which is more than grammatically different from other words we need an extra definition. But for all (different) DVs which belong to the same DS we need only one definition - the definition of the DS. This definition is usually also more precise than the definition of a word, and internationally valid. DSs can be created by web users according to their domains of interest, so it is a general approach. There can be much more different DVs (even in only one DS) than there are different words. Therefore DV based description and search has higher range and resolution than word based description and search.

As described in section 3, DV based description additionally provides information about similarity relations of resources.

### 8.2 Precise information exchange on the web

An important motivation for this approach is improvement of the availability (this means also searchability) of precise information on the web. For description of reality usually words of language are used, but they categorize the original quantitative features of reality. At this often interesting information gets lost. Even if someone wants to provide precise information and explicitly adds quantitative information, e.g. as numbers combined with text, up to now usually only the words of the text are searchable, not the numeric quantitative information. In many cases just this precise quantitative information is interesting for the readers. So it is reasonable to combine numbers with unique identifiers, that they are machine readable and searchable. This is done in a DV: It contains the DSI (URL of the DS), which together with the DI (Dimensions Identifier, see 7.2.1) or the position in a semicolon separated list uniquely identifies the numeric value of every given dimension.

The DS definition and its identifiers are also a guide for providers (writers) of numeric information. Often important numeric data are missing on the web, because the writer does not know well the expectations and interests of the reader. The DS definition shows the quantitative data, which are in a certain domain interesting for the readers. So it serves in this domain as standardized and expandable interface for exchange of precise numeric information between writer and reader. Later it is possible to provide to dimensions frequencies of usage.

### 8.2.1 From DSs derived evaluation DSs

Correctness is precondition of precision, and is it possible for interested companies (e.g. search engines) to give their users the possibility to evaluate every original DV by an own "evaluation DV". For this from original DSs automatically "evaluation DSs" can be derived, which contain "evaluation dimensions" for every (unbranched) value of the original. These evaluation dimensions can branch again as DSs (Fig. 2) into values, e.g. "correct value", "|value| / |correct value|" (automatically calculated with upper limit), "subjective grading of precision" (0...15), "subjective grading of reliability" (0...15) etc. Every "evaluation DV" of such an "evaluation DS" contains the URL (7.2) of the original evaluated DV as searchable dimension (e.g. for statistics) and can be created by every registered user. Because an evaluation DS is again a DS, there can be also an evaluation DS of an evaluation DS (if there is interest).

### 8.3 Storage of DS definitions open on the web, storage of DVs on the web and locally

The implementation shows that realization of user defined DSs and DVs is also possible in a local online database with Numeric Search engine. The Vector Location (VL, see 4.1) cannot be used in a local database, but the DVs can get a hyperlink to a location on the web, which they describe. So it would be technically feasible to realize this Numeric Search also in a local database.

This is better than nothing, especially if certain data cannot be published openly on the web, e.g. patient records with detailed quantitative data about medical findings, treatment and treatment outcome. Even if these data are not published, Numeric Search and anonymous statistical output of the results (e.g. average values of dimensions) can be used for decision support.

But it requires considerable effort that a local (proprietary) database is used internationally. There is relevant probability that several competing databases arise, so that Numeric Search is no more internationally complete but restricted to one of many proprietary databases (information silos). Therefore we recommend introduction of a web standard for worldwide valid DSs whose definitions should be always placed open on the web. Their elements (DVs) can be also private (if necessary in local databases), but usually DVs should be also placed as open data on the web, because this is most informative. Besides making quantitative data searchable, DVs (identified numbers, see 7.4) can be also used for interoperable exchange of quantitative data in machine readable form.

### 8.4 Reliability of DS definitions

The numeric data in a DV are only meaningful together with their definition in the DS. Therefore every dimension definition of a DS must be stable. When in the course of time a dimension turns out to be no more recommendable for new data, it can be marked as "deprecated", and an explanation can be added. For this purpose we recommended usage of a keycomment pair (see section 6.7) with a fixed and a changeable part. A checksum can be calculated from every fixed part of a dimension definition, which can be also integrated in DVs. So change of a definition would be detectable. But recovery of a dimension definition is only possible from a backup copy. To guarantee reliability, DS definitions can be stored in reliable (official) web sites, which are open for read and which allow expansions and changes of DS definitions only in non-fixed parts. This signals to providers of numeric data (DVs) that the DS definitions are stable. Additionally Numeric Search engines can create backup copies of (frequently used) DS definitions on the web and mark changes of DS-definitions.

## 8.5 Definition of DSs by web users

To cover the range of topics on the web, those who create the web should be also able to routinely define DSs, so that they can make useful definitions about all topics which are of common interest. Appropriate Software can considerably facilitate generation of DS definitions and DVs, and interpret these. Web users can define DSs according to their expertise and domain of interest. They can define the numeric content and also the meaning of the hyperlinks (Fig. 7) of keycomments of DVs. The owner of a DS can for example define that the first hyperlink of the keycomment (see 6.1) of every contained DV points to a specific dataset (e.g. a picture, song, data generated by software of the DS owner) and the numeric content is a searchable specific feature extraction of this dataset. If the feature extraction is appropriate, the DV makes the dataset available to DS specific similarity search by all Numeric Search engines.

Different formats are possible as concrete syntax for DS definitions and DVs. Because the length of the usual formats can obstruct overview and performance we prefer an as short as possible form at least for DVs. The main point is that the numbers in the DVs are identified (see 7.4) so that they can be associated to the dimension of a DS. This can be done in varying environment (see e.g. 8.9).

## 8.6 Equivalent definitions, usage of sameAs

Before defining a new DS or DS dimension it is important to check existing definitions. If there is already a definition which completely covers the intended purpose, a new definition is not necessary. If nevertheless a new definition is wished, e.g. because the new definition should contain a link to a certain website, it is possible to create a new equivalent definition without unnecessary adverse effect, if a link to the equivalent definition is inserted using the "sameAs" statement (like owl:sameAs [W3C 2004]). This shows to the search engine, that by "sameAs" connected identifiers (URLs of DSs or DS dimensions) refer to the same thing and (generation of a synchronized index 6.5 and) search could be done (instead only over one identifier) more completely over all connected identifiers together.

*Recommendations for creators of definitions:* To find equivalent definitions, existing definitions can be checked e.g. by specific text search within (selectable parts of) DS definitions. If there is an existing DS (let's call it "DS-owned") and few dimensions are missing, this can be told to the DS-owner. The owner can add missing definitions to keep the DS attractive. If nevertheless a new definition is necessary, instead of defining a DS completely new, it is better to include (see 5.2) suitable DS definitions (also DS dimension definitions, see 7.1.1 - e.g. it is possible to include the dimensions of "DS-owned"). If instead of inclusion equivalent (re)definition is preferred, equivalent definitions should be linked together using the "sameAs" statement. Included or (via "sameAs") connected definitions get higher search frequency than isolated definitions.

*Recommendations for providers of DVs:* To find the most relevant suitable DS for quantitative data, search engines can be asked using specific text search within (keywords or comments of) DS definitions to get a list of DSs which touch a certain Domain or topic. The list can be ordered e.g.

by the size of the DSs (the count of contained DVs) or search frequency. This can be used to find the most relevant DS definition. A check of their definitions can help to find the best fitting DS.

## 8.7 Nested DS definitions

According to section 5.2 every DS definition combines dimensions which can represent not only numeric values but also again a (part of a) DS definition. This possibility allows complex expansions and generates additional similarity relations. So e.g. an included DS definition with URL http://example.org/DS1 defines additional similarity relations to DVs of the DS http://example.org/DS1 and DVs of all other DSs which include the DS http://example.org/DS1. Because this included DS definition again can be nested, ontological structures of DS definitions with high dimensionality are possible. Such structured definitions of complex DSs are meaningful due to several reasons. Besides the definition of additional connections (similarity relations) and reuse of existing definitions they can provide a subdivided and structured representation of the domain. So Domain Spaces can also represent user defined conceptual spaces which were proposed by Gärdenfors [Gaerdenfors 2000; 2004].

A special case needs attention:

### 8.7.1 Infinite nesting level

There is the possibility of a circular definition in a DS: A DS definition (e.g. http://example.org/person) can include as dimension again (DS definitions which include) the same DS definition (http://example.org/person as friend). Obviously a DV of such a DS can provide content (numerical values) only up to a limited count of circular expansions. There is, however, the possibility (see 7.2.1) to include into a DV as dimension instance the URL of another DV instead of including directly the numerical values. Expansion of the DV is in this case possible until the chain of included (URLs of) DVs stops or (circularly) goes back into itself. For search such deep expansions are usually not meaningful. A search engine can at index calculation limit the count of circular expansions, or generally limit the nesting level which is traced.

## 8.8 High dimensional DSs

Extensions and nesting of DSs definitions can soon lead to high dimensional DSs. High dimensional similarity search in these spaces tends to become inefficient [Aggarwal et al. 2001], but the dimensions can serve as large set of possibilities for low dimensional search, and as basis for exchanging quantitative information. High dimensional DS definitions can serve as container of many one dimensional DS definitions. These are accessible using the URLs of the dimensions, see 7.1.1. So it is possible to reuse and combine dimensions in new nested DSs.

## 8.9 Application in Linked Open Data (LOD)

All RDF triples [World Wide Web Consortium 2014] which contain short text (URLs or generally URIs) as subject, predicate and object can be represented as elements of a single DS with three "long" dimensions, see 6.7 . The efficiency of DVs could be, however, better used by 6.9.
Generally the triple pattern is only one possibility to describe a relation between dimensions. A DS definition contains an ordered set of dimensions (Fig. 1), so it can also contain the definition of a certain relation between the dimensions 7.1.

An efficient and simple way to realize connected data (connecting data is motivation of the LOD cloud [Bizer et al. 2009]) is to define appropriate DSs for the interesting numeric data, and to identify these (e.g. using the form in 7.4), and to group the data in a DV together. Then additionally to the connections via hyperlinks these numeric data (of a DV) define bidirectional similarity relations ("numeric links") to the (numeric data of) other DVs of the same DS (Fig. 4) and (in case of a nested DS definition) to the DVs of included DSs. Identifying numbers (as components of DVs) provides these connections generally between numeric data on the web.

### 8.10 Constructing DSs from frequently used dimensions (with medical example)

Building high dimensional DSs can be reasonable e.g. for providing patterns of commonly used dimensions within a large domain, for later construction of derived DSs and for special conversation within this large domain. An example: Medicine deals with a lot of quantitative data which can be derived from diagnostic measurements on the patient, treatment data, result data, derived data (also human generated medical classifications) etc.. These quantitative data can be combined as dimensions of large expandable DSs which serve as standardized initial container. Then statistical data can be obtained by observing, which dimensions (quantitative data) of these containers are used by physicians (users) in case of which situation, e.g. ICD diagnosis [WHO 2014] (this can be expanded to all helpful sets of distinguishable situations, also again to distinguishable measurement results). Then we get for different situations (e.g. diagnoses) frequencies of used dimensions (quantitative data, e.g. measurements). This information can be used again by physicians (users). If the information about frequently used quantitative data in a certain situation is taken into account, there is less probability that important data (measurements) are neglected. Additionally the statistics shows natural connections between different situations: Situations (diagnoses) with the same most frequently used quantitative data (dimensions) can be grouped together. Later DSs can be defined from these dimensions and associated to these groups of situations (diagnoses) - for better clearness. Because these DSs are derived from natural frequencies of used quantitative data (dimensions), they are less dependent on the initially used nomenclature for certain situations (human created names of diagnoses). They depend on the original natural situation, and they can serve as interface for exchange of searchable objective quantitative data in this situation.

### 8.11 Decision support (with medical example)

Decision support (concerning measurable reality) is primary motivation for information processing. Due to the basal importance of this topic we provide a medical example. For decision support the dimensions of a Domain Space can be subdivided into 3 parts:

{1} **Preconditions** (In Medicine: Findings)
{2} **Decision** (In Medicine: Treatment)
{3} **Result**

In daily practice much valuable information arises in medicine, especially information about the results after this or that treatment (decision). Up to now most of this information gets lost after some time and it is no more available for the community. To make it available we need to store a description of 3 states or procedures: {1} The precondition (e.g. medical finding), {2} the decision (e.g. selection of treatment) and {3} the situation (result) enough time afterwards. (The dimensionality of the result {3} can be extended retrospectively, e.g. if additional consequences of a treatment become known.) The description of {1}{2}{3} should be reproducible, precise and searchable in sufficient resolution. DVs fulfill the requirements. If they contain respectively the sequences {1}{2}{3}, users can get decision support by searching descriptions of {1} and/or varying {2} and looking for the result {3}. Because this deals with quantitative data, immediate statistics "near" the searched description are possible, to find the variant {2} which leads to the best result {3}.

Medical example:

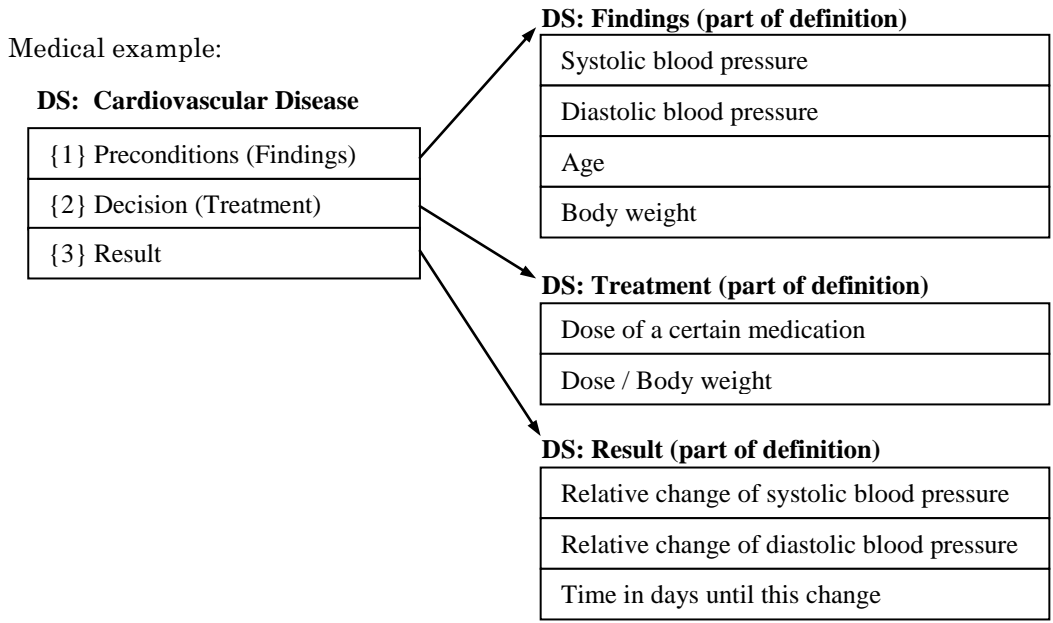

Fig. 20. Exemplary DS definition to evaluate the relative change of blood pressure{3} in dependence of medication {2} in case of a certain individual situation {1}. Only a part of relevant dimensions is shown.

So it is e.g. possible to check in case of individual preconditions {1} (Fig. 20) immediate statistics of result dimensions {3} (e.g. change of blood pressure in Fig. 20) in dependence of selected ranges of influencing therapeutic dimensions (e.g. decisions {2} like dose of medication in Fig. 20).

It can be appropriate (medical data are an example for this) to calculate new derived DVs from existing DVs. So in medicine the chronological process of health is important, therefore it is appropriate to calculate from the original DVs of a certain patient derived (often higher dimensional) DVs which summarize measurements and changes at a certain interval of time. For this the original DVs need to contain a unique pseudonym (e.g. a number) of the patient.

Generally the practical benefit of decision support depends on the completeness of available data. In case of simple situations which depend on few dimensions, e.g. certain technical data, it is easier to provide enough data for decision support than in case of situations which depend on many dimensions. Because real life situations contain (of course) too many dimensions for a complete acquisition and description using available hardware, approximation is necessary using few dimensions within a restricted domain of interest. Finding the most relevant influencing dimensions for a certain domain of interest (a DS) can be a demanding process. The completeness of these dimensions can be estimated. It is the better, the smaller in case of controlled dimensions {1}{2} the standard deviations of the dependent result dimensions {3} are. Statistics can be used to detect dependencies. *Statistical results are the more reliable, the larger the database is.* This is one of many arguments for *worldwide* DSs.

### 8.11.1 Efficient definition of domain ontologies

Many arguments underline the need for domain ontologies, e.g. in medicine [Haas 2005]. There are already examples for ontologies which are large (e.g. Snomed CT) but nevertheless not fine enough for decision support without refinement by quantitative spaces like DSs (see Fig. 20). Therefore instead of defining non-quantitative ontologies it is more efficient to define ontologies from the beginning using DSs at the vertices (instead of points resp. words). At this we can define the DSs that they cover a maximal set of possibilities using partially the same dimensions, to minimize the count of necessary vertices and definitions and to maximize the connections.

Also retrospectively definitions of DS dimensions can be connected via their URLs in multiple ways, e.g. using algebraic expressions, attributes (e.g. "sameAs", "extents").

### 8.12 Decision support in jurisdiction

DVs can make legislative text more precise. Similar like for description of medical decisions DVs can be also used for description of judgments and (internationally) large searchable web collections of judgments can be built. So it would be possible for judges to compare existing cases to past cases in the collections more precisely and to check past judgments. This could help jurisdiction towards better reproducibility and precision.

Like in medicine, also in jurisdiction there are far reaching possibilities, also concerning usage of support by algorithms which use DVs as input and output. It would exceed the scope of this paper to deepen this here.

### 8.13 Graphical representation of DVs

A DV can represent information more precisely than a word of language. DVs can be embedded into HTML, after click on it the definition of every used dimension can be shown. There are many possibilities for graphical representation of the quantitative data. For example the value of every dimension can be represented graphically in comparison with the values of other DVs of this DS, together with graphical representation of statistical information (e.g. percentile, count of standard deviations from the mean etc.). A further possibility is e.g. graphical representation of comparisons of DVs.

### 8.14 Scientific communication, impact factor of published data

Scientists who make research within a special domain can define DSs of their specialty whose dimensions represent measurements and statistical results which are interesting in their group. By looking at DS definitions and statistics scientists get information about searched data already before deciding for a measurement. Appropriately defined DSs allow scientists to search and extract just the data which they need.

Obviously DSs provide efficient means for storing, and exchanging scientific data in searchable form. Scientists can publish (*not only articles but also*) quantitative *data* in DSs of their specialty. Authors of scientific articles can make citations of such published quantitative *data* (and optionally estimate, how much percent these data contributed to the results of the article). This can form the basis of an impact factor of published *data*.

The frequently used Journal Impact Factor (JIF) is only one possibility and only one dimension of a possible multidimensional quantification of scientific impact which could be mapped into a Domain Space.

### 8.15 Comparison to the Vector Space Model in information retrieval

Up to now the Vector Space Model in information retrieval is mainly used to compare text documents [Wikipedia: Vector space model 2014]. Vectors which represent text documents represent quantitative data of these texts (e.g. frequencies of keywords), so these can be seen a special case of Domain Vectors.

### 8.16 To the efficiency of the concept

According to [1] every DV (element of a DS) is represented by a
> URL plus a (ordered) *sequence of numbers*.

By "URL" we mean an (internationally) unique resource locator (literally speaking). It can be e.g. the uniform resource locator defined in [W3C URL], it can be (later) also e.g. a unique number (which could be more efficient) or a locally defined substitute.

The compact form "URL plus sequence of numbers" is sufficient for definition of a DV due to the well defined *sequence of dimensions* and the URL of a DS definition. It is also necessary for uniform definition of information (the URL determines the domain and the numbers the selection within the domain, see [1] ). So we can avoid redundancy.

Because one DS definition defines every DV (element) of the DS (not only one word), a huge extension of the vocabulary is possible.

### 8.17 "Quantified" text with DVs as words

Important part of a DV [7.2] is the URL of the (definition of the) containing DS. We can arrange, that in case of certain URL names optionally a certain part is shown, e.g. if the URL has the form part1--part2--part3 then only part2 is shown. Part2 can be a usual word of language, e.g. "go". Then the URL "part1--go--part3" could locate a DS definition with the first dimension "speed in km/h", and every DV of this DS is shown by the word "go", formatted as DV, and click on it shows all given quantities, e.g. "5 = speed in km/h". So DVs could become part of "quantified" text, which contains "quantified" words which provide additional quantitative information.

Additionally the DS Definitions at URLs with form part1--part2--part3 can contain translations translatedpart2 of part2 in other languages, where at the URLs part1--translatedpart2--part3 a link to part1--part2--part3 is given. If wished, after input of part2 text software can automatically create DVs with URL part1--part2--part3, click on these DVs could show besides possible quantitative data also translations.

(Already today a text viewer could generate from a word "part2" a link to a URL part1--part2--part3, where translations of part2 could be placed.)

### 8.18 Motivation for owners of DSs

Motivation for the owners of DSs definitions is better communication within their domain of interest, and the possibility to expand the DSs subsequently by additional interesting dimensions. Moreover owners can modify the changeable part of DS definitions (see 6.7) and provide links e.g. to their web pages.

Patents on DS definitions, however, should not be possible. The reason for this is given in the following remark:

### 8.19 Language is not patentable

There must be open access to DS definitions so that they can be valid worldwide. The proposed standard for worldwide valid DSs allows to include (reuse) DS definitions in new definitions and to extend definitions a posteriori. The approach is designed for free and efficient usage of data on the web. Patents on DS definitions would contradict this purpose. Moreover: DSs define precise quantitative descriptions worldwide. This can be seen as an extension of language (see chapter 1). Patents on (parts of) language are not possible. Therefore patents (or "copyright" etc.) on DS definitions are contraproductive and should be impossible. Also free content (DVs) is desirable.

## 9. CONCLUSION

DSs can be defined by web users according to their domains of interest. Their elements, the DVs, identify quantitative data and so make these data machine readable and searchable. Nested DS definitions realize hierarchical ontologic structures, where common identifiers (URLs) of dimensions provide connections between different DSs and enable general similarity search of quantitative data on the web. DV based description and search (Numeric Search) can become an important addition to usual word based description and search on the web. Therefore the introduction of a web standard for worldwide valid DS definitions and DVs is recommendable.

### REFERENCES

Aggarwal, C. Hinneburg, and A. Keim, D. 2001. On the Surprising Behavior of Distance Metrics in High Dimensional Space. ICDT 2001, Vol. 1973, 420-434. DOI:http://dx.doi.org/10.1007/3-540-44503-x_27

Andoni, A. and Indyk, P. 2006. Near-optimal hashing algorithms for approximate nearest neighbor in high dimensions. Communications of the ACM, vol. 51, no. 1, 2008, p. 117-122. http://mags.acm.org/communications/200801/#pg119

Berners-Lee, T., .Hendler, J., and Lassila, O. 2001. "The semantic web". Scientific American 284.5, 28-37. http://www.cs.umd.edu/~golbeck/LBSC690/SemanticWeb.html

Berners-Lee, T., Fielding, R., & Masinter, L. 1998. Uniform resource identifiers (URI): generic syntax. Retrieved April 1, http://www.hjp.at/doc/rfc/rfc3986.html

Bizer, C., Heath, T., & Berners-Lee, T. 2009. Linked Data - The Story So Far. International Journal on Semantic Web and Information Systems (IJSWIS), 5(3), 1-22. http://dx.doi.org/10.4018/jswis.2009081901

Black M. 1963. Reasoning with Loose Concepts. *Dialogue*, 2, 1-12. http://dx.doi.org/10.1017/S001221730004083X

Fontoura M., Lempel R., Qi R., Zien J. 2006, Inverted Index Support for Numeric Search, *Internet Mathematics*, 3(2), 153-185. http://projecteuclid.org/euclid.im/1204906137.

Gärdenfors, P. 2000. Conceptual Spaces: The Geometry of Thought. MIT Press, Cambridge, MA.

Gärdenfors, P. 2004. How to make the semantic web more semantic. In A.C. Vieu and L. Varzi, editors, Formal Ontology in Information Systems, pages 19–36. IOS Press, 2004. http://yaxu.org/tmp/Gardenfors04.pdf

Gionis A., Indyk P., Motwani R. Similarity Search in High Dimensions via Hashing, Proceedings of the 25th Very Large Database (VLDB) Conference, 1999, p. 518-529. http://dl.acm.org/citation.cfm?id=671516

Grant, E., Hegde, C., Indyk, P. Nearly optimal linear embeddings into very low dimensions, Global Conference on Signal and Information Processing (GlobalSIP), 2013 IEEE, p. 973-976. http://ieeexplore.ieee.org/xpl/articleDetails.jsp?arnumber=6737055&tag=1

Guyon, I., Gunn, S., Nikravesh, M., Zadeh, L., ed. 2006. Feature Extraction, Foundations and Applications. Studies in Fuzziness and Soft Computing, Springer, Berlin, Heidelberg.

Haas, P. 2005. Medizinische Informationssysteme und Elektronische Krankenakten. Springer, Berlin, Heidelberg.

Haveliwalay T. H., Gionisz A., Kleinx D., Indyk P. 2002. Evaluating Strategies for Similarity Search on the Web. In Proc. WWW 2002, 432-442, ACM Digital Library.

Heinz, S. and Zobel, J. 2003, Efficient single-pass index construction for text databases. J. Am. Soc. Inf. Sci. 54, 713–729. http://dx.doi.org/10.1002/asi.10268

Holsgrove, G. and Elzubeir, M. 1998. Imprecise terms in UK medical multiple-choice questions: what examiners think they mean. *Medical Education* 32, 343–350. http://dx.doi.org/10.1046/j.1365-2923.1998.00203.x

Indyk P., Motwani R. 1998. Approximate Nearest Neighbors: Towards Removing the Curse of Dimensionality, Proceedings of 30th Symposium on Theory of Computing, p. 604-613. http://dl.acm.org/citation.cfm?id=276876

Indyk P. 1999. Dimensionality Reduction Techniques for Proximity Problems, Stanford University. (650-723-4532). http://people.csail.mit.edu/indyk/proxy99.ps

Klyne, G., & Carroll, J. J. 2004. Resource description framework (RDF): Concepts and abstract syntax. http://www.w3.org/TR/2004/REC-rdf-concepts-20040210/

Kolmogorov, A. N. 1968. Three approaches to the quantitative definition of information. Problems of information transmission, 1(1), 1-7.

Kriegel, H.P., Kröger, P., Renz, M., Schubert, M. 2010. Metric spaces in data mining: applications to clustering. *SIGSPATIAL* Special Volume 2 Issue 2 (July 2010), 36-39. http://dl.acm.org/citation.cfm?doid=1862413.1862423.

Limaye, G., Sarawagi, S. and Chakrabarti, S. Annotating and searching web tables using entities, types and relationships. In VLDB, 2010. http://vldb.org/pvldb/vldb2010/papers/R118.pdf

MATH41002: LINEAR ANALYSIS, viewed May 2014, Minkowski inequality http://www.maths.manchester.ac.uk/~nikita/31002/minkowski.pdf

McGuinness, D. L., & Van Harmelen, F. 2004. OWL web ontology language overview. *W3C recommendation*, *10*(10)

Nakao M. A., Axelrod S. Numbers are better than words. 1983.Verbal specifications of frequency have no place in medicine. Am J Med 1983, 74, 1061–1065. http://dx.doi.org/10.1016/0002-9343(83)90819-7

Movable Type Scripts, viewed May 2014. Calculate distance, bearing and more between Latitude/Longitude points, http://www.movable-type.co.uk/scripts/latlong.html

Orthuber, W., Fiedler, G., Kattan, M., Sommer, T., Fischer-Brandies, H. 2008. Design of a global medical database which is searchable by human diagnostic patterns. *The Open Medical Informatics Journal* 2, 21-32. http://www.ncbi.nlm.nih.gov/pmc/articles/PMC2666959/

Orthuber, W., Dietze, S. 2010. Towards Standardized Vectorial Resource Descriptors on the Web. 2010. In *Lecture Notes in Informatics*, INFORMATIK 2010. Service Science - Neue Perspektiven für die Informatik. Band 2 P-176, 453-458 http://subs.emis.de/LNI/Proceedings/Proceedings176/469.pdf

Orthuber, W., Papavramidis E. 2010. Standardized Vectorial Representation of Medical Data in Patient Records, *Medical and Care Compunetics* 6, 153-166. http://science.icmcc.org/2010/06/20/standardized-vectorial-representation-of-medical-data-in-patient-records/

Orthuber, W. NumericSearch - online implementation of vectorial description and search. 2012. (Online since July 2012) http://NumericSearch.com/

Orthuber, W. General approach to similarity search of resources with numeric features on the web. 2013. Open Data on the Web. http://www.w3.org/2013/04/odw/papers#al1

Orthuber, W. How to make quantitative data on the web searchable and interoperable part of the common vocabulary. in GI-Jahrestagung 2015.

Pimplikar, R., Sarawagi, S. Answering table queries on the web using column keywords. In Proc. of the 38th Int'l Conference on Very Large Databases (VLDB), 2012. http://vldb.org/pvldb/vol5/p908_rakeshpimplikar_vldb2012.pdf

Sarawagi, S., Chakrabarti, S. 2014, Open-domain quantity queries on web tables: Annotation, response, and consensus models. In ACM SIGKDD, 2014, 711-720. http://dx.doi.org/10.1145/2623330.2623749

Schema.org, viewed May 2014. Getting started with schema.org, in http://schema.org/docs/gs.html

W3C 2004, OWL Web Ontology Language, W3C Recommendation 10 February 2004, owl:sameAs, in http://www.w3.org/TR/owl-ref/#sameAs-def

W3C 2014 Uniform Resource Locators (URL)  https://www.w3.org/TR/url

WHATWG community, HTML Living Standard, viewed May 2014. Microdata, in http://www.whatwg.org/specs/web-apps/current-work/multipage/microdata.html#microdata

Wikipedia, viewed March 2015. Big Data, in http://en.wikipedia.org/wiki/Big_data

Wikipedia, viewed May 2014. Domain model, in http://en.wikipedia.org/wiki/Domain_model

Wikipedia, viewed May 2014. Domain ontologies and upper ontologies, in http://en.wikipedia.org/wiki/Ontology_(information_science)

Wikipedia, viewed May 2014. Feature extraction, in http://en.wikipedia.org/wiki/Feature_extraction

Wikipedia, viewed May 2014. Linear subspace, in http://en.wikipedia.org/wiki/Linear_subspace

Wikipedia, viewed May 2014. Metric Space, http://en.wikipedia.org/wiki/Metric_space

Wikipedia, viewed May 2014. Minkowski Distance, http://en.wikipedia.org/wiki/Minkowski_distance

Wikipedia, viewed May 2014. Quantity, in http://en.wikipedia.org/wiki/Quantity

Wikipedia viewed May 2014, Resource Description Framework (RDF), http://de.wikipedia.org/wiki/Resource_Description_Framework

Wikipedia, viewed May 2014. Vector Space Model, in http://en.wikipedia.org/wiki/Vector_space_model

World Health Organization, viewed May 2014. International Classification of Diseases (ICD), in http://www.who.int/classifications/icd/en/

World Wide Web Consortium. 2014. RDF 1.1 Concepts and Abstract Syntax. http://www.w3.org/TR/rdf11-concepts/

Zezula, P., Amato, G., Dohnal, V., Batko, M. Similarity Search. The Metric Space Approach. 2005. *Series: Advances in Database Systems*, Vol. 32., Springer, Berlin, Heidelberg.